\documentclass[graphics,floatfix,nofootinbib,notitlepage,
tightenlines,nobibnotes,aps, aps,twocolumn,superscriptaddress]{revtex4-1}
\usepackage{amsmath, graphicx,multirow, verbatim, psfrag}
\usepackage{amssymb, amsthm}
\usepackage{floatpag}
\usepackage[margin=18mm]{geometry}
\usepackage{fancyhdr}
\usepackage{wrapfig}
\fancypagestyle{newstyle}{
\fancyhf{} % clear all header and footer fields
\fancyfoot[R]{\thepage} % except the center

}
\theoremstyle{remark}

\makeatletter
\@ifundefined{textcolor}{}
{%
 \definecolor{BLACK}{gray}{0}
 \definecolor{WHITE}{gray}{1}
 \definecolor{RED}{rgb}{1,0,0}
 \definecolor{GREEN}{rgb}{0,0.5,0}
 \definecolor{BLUE}{rgb}{0,0,1}
 \definecolor{CYAN}{cmyk}{1,0,0,0}
 \definecolor{MAGENTA}{cmyk}{0,1,0,0}
 \definecolor{YELLOW}{cmyk}{0,0,1,0}
 }
  \usepackage[usenames,dvipsnames]{xcolor} %\usepackage[dvips]{color}

\usepackage{subfig,subfloat,float}
  
\begin{document}
\title{Weak transcription factor clustering at binding sites can facilitate information transfer from molecular signals}
\author{Tamara Mijatovi\'c}
\author{Aim\'ee R. Kok}
\author{Merlijn Br\"uggen}
\author{Jos W. Zwanikken}
\author{Marianne Bauer}
\email{m.s.bauer@tudelft.nl}
\affiliation{
Bionanoscience Department, Kavli Institute of Nanoscience, Delft University of Technology, van der Maasweg 9, 2629 Delft, The Netherlands
}

\begin{abstract}
Transcription factor concentrations provide signals to cells that allow them to regulate gene expression to make correct cell fate decisions.
 Calculations for noise bounds in gene regulation suggest that clustering or cooperative binding of transcription factors  decreases signal-to-noise ratios at binding sites. However, clustering of transcription factor molecules around binding sites is frequently observed. 
 We develop two complementary  models for clustering transcription factors at binding site sensors that allow us to study information transfer from a signal, the morphogen Bicoid, to a variable relevant to development, namely future cell fates. We find that weak cooperativity or clustering can allow for maximal information transfer, especially about the relevant variable. The timescale of measurement is crucial for predicting the optimal clustering strength: for short measurements, clustering allows for the implementation of a switch, while for long measurements, weak clustering allows the sensor to access maximal developmental information provided in a nonlinear signal. Finally, we find that clustering 
not only facilitates information maximization about the relevant variable, but also can allow the binding site sensors to achieve optimality in a related optimization goal, the information bottleneck (IB) bound. While the measurement time restricts the region on the information plane that is accessible, changes in clustering in conjunction with changes in the binding energy can shift the binding site along the optimal bound, and towards an optimal trade-off between obtaining information about the signal and obtaining relevant information.
\end{abstract}
\maketitle

Cells differentiate and develop in response to chemical signals which convey information.
It has been proposed that these cellular responses follow principles of information maximization -- in  development and gene regulation, but also in neuronal signaling or chemotaxis \cite{bialekbook, barlow, Palmer, koshland, mattingly2021, Tjalma, reddy, Tkacik_2008, selimkhanov2014accurate, cheong2011information}.  
Early fly development is one of the canonical examples for precise information flow \cite{Tkacik_2008, Gregor2007a, Dubuis, Desponds2020, Baueretal, sokolowski25}. There, a few maternal morphogens, including the transcription factor Bicoid, regulate the development of cells along the head-to-tail body axis of the embryo \cite{Nusslein-Volhard_1980, driever1988gradient, Gregor2007a}. This regulation occurs via a small network of genes, including four of the gap genes \cite{Nusslein-Volhard_1980, Jaeger_2011}, with an expression profile precise to \textit{ca.}  0.1\% of the embryonal length \cite{Dubuis}.
Many parts of cell fate development, such as gap gene patterns and their regulation of cell fates, have been shown to be consistent with principles of information optimization with respect to cell fates \cite{Dubuis, Petkova, Baueretal, sokolowski25, Nikolic}.

On the molecular level, this information transfer is facilitated by binding of transcription factors to binding sites in the regulatory regions of the genome,  including promoter or enhancer regions \cite{furlonglevine_18}. While the information optimality of binding site sensors is difficult to assess due to the requirement of knowledge of multiple parameters, precise experiments on promoter statistics  and binding site occupations are increasingly pushing this question to an empirical forefront \cite{ChenLevoZoller, fallacaro2025, Munshi_2024,Mir_2018}.

Yet, recent experiments on transcription factor behaviour around binding sites might pose a challenge to these information-maximization principles.
Transcription factors 
show clustered distributions inside cells or around these binding sites \cite{ Mir,Mir_2018, Dufourt_2018,Munshi_2024}. 
This clustering is associated with reduced information flow, since earlier theoretical calculations based on the Berg-Purcell bound \cite{Berg-Purcell, BialekSetayeshgar, tenWolde, Kaizu} found that the signal-to-noise ratio at binding sites decreases when binding occurs cooperatively \cite{BialekSetayeshgar2} or via local clustering \cite{SkogeWingreen, SkogeII}.  
 We use the phrase `clustering' here to remain agnostic of the precise mechanism that leads to a positive affinity between proteins \cite{McSwiggen_2019}, as local clusters can indeed occur from a variety of mechanisms: transcription factors can cluster, for example, by binding cooperatively at binding sites \cite{Lebrecht}, or by forming larger clusters around the binding site region, potentially as a consequence of binding or biomolecular condensation 
\cite{Hyman_2014,Shin_2017, Hnisz_2017,Sabari_2018, McSwiggen_2019, Hannon_2024}.

Here, we address this apparent conflict of information-maximization principles in early fly development by investating in detail how transcription factor clustering at binding sites affects the information transfer. We shift the focus from sensing errors and signal-to-noise ratios \cite{BialekSetayeshgar2, SkogeWingreen, SkogeII, malaguti2021theory, mora2019physical} to the information that the binding site captures about the signal and conveys about possible downstream cell fates, the variable of functional relevance. We observe that this shift in focus resolves the apparent conflict. With two complementary models, we find that weak clustering is consistent with optimizing at least this variable of relevance. In addition, transcription factor clustering can help achieve optimality consistent with the information bottleneck goal \cite{tpb00}, and can tune binding sites toward obtaining relevant information with a comparatively noisy measurement. We find further that the timescale in the biological system, i.e. the time during which transcription factor occupations are measured by the binding site, is crucial for predicting the optimal amount of clustering, also for the information about the signal.  

Our paper is structured as follows: In section \ref{sec:I}, we introduce the problem of information transfer, here based on only one of the maternal signals in fly embryo development through a single binding site sensor, our timescale estimates and the calculation procedure. In section \ref{sec:Hill}, we guide expectations  
 with a model established in the context of cooperative binding, where the occupation follows a Hill-function. 
This model suggests that weak cooperativity improves information transfer.  
In section \ref{sec:mechanism}, we develop a more 
realistic 
 model of molecules clustering at and around a single binding site sensor, 
inspired by Ising models \cite{SkogeWingreen, morin2022sequence, Hillenbrand}, that goes beyond cooperative interactions. With this model, we find that clustering affects information quantities differently, but that weak clustering is beneficial for short and barely affects relevant information for long timescales. Finally, we show in section \ref{sec:ib} that with both models, different parts on the optimal information bottleneck bound are achieved by different clustering strengths for different timescales, but that tuning clustering in conjunction with binding site strength allows binding site sensor to tune along the optimal IB-bound. In section \ref{sec:dis}, we summarize and discuss future improvements of our work, also in the context of new measurements that may allow to assess information optimality of binding site sensors based on experimental data, either directly through mRNA production \cite{ChenLevoZoller} or measurements of transcription factor occupations at binding sites \cite{Bothma, Munshi_2024, fallacaro2025}.

%%or: 

\section{Information Flow through binding site sensors \label{sec:I}}

\begin{figure}
\centering
\includegraphics[width=\columnwidth]{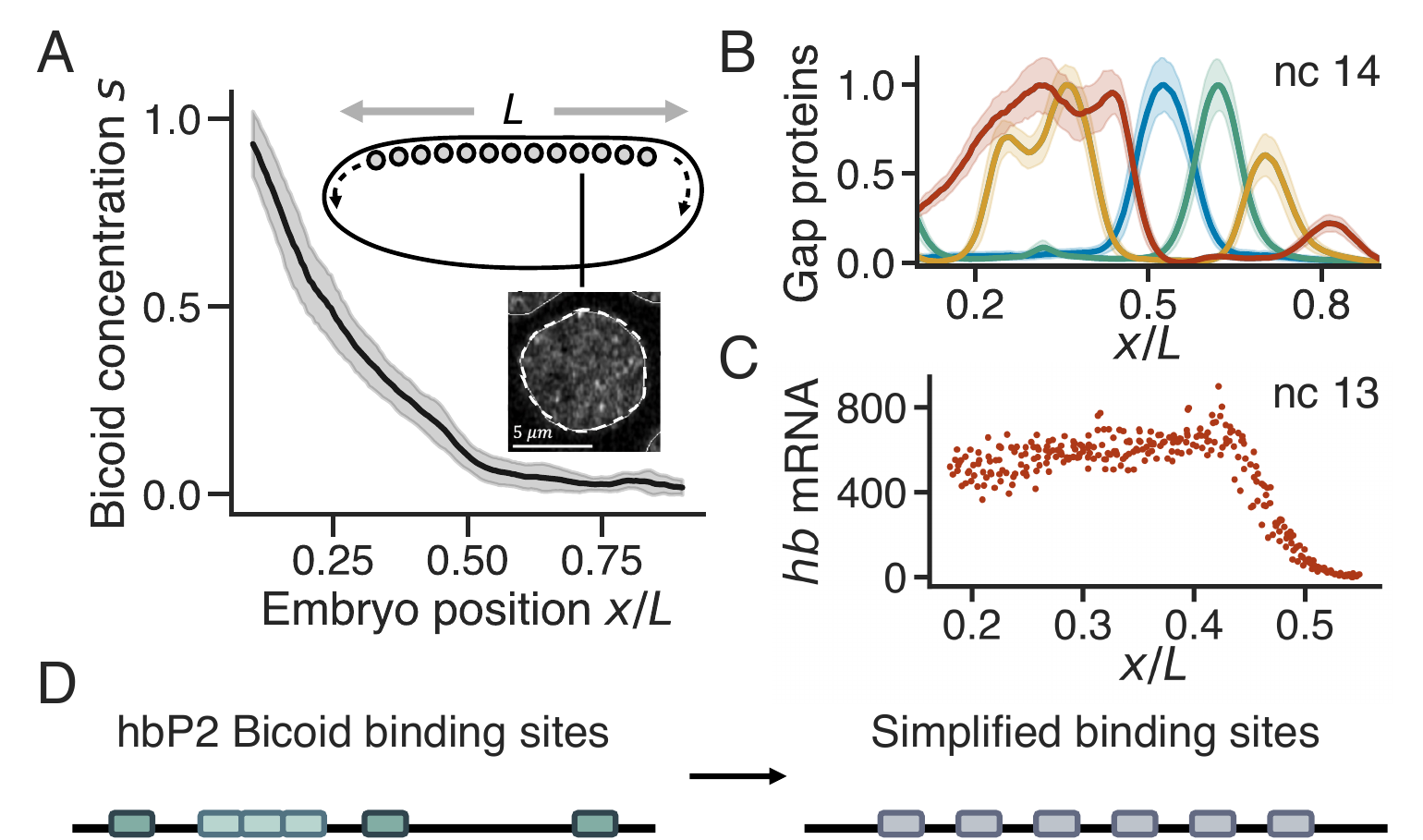}
\caption{ Signal processing in the fly embryo. A) The normalized maternal Bicoid gradient \cite{Gregor2007a} along the embryonal head-to-tail axis, which regulates the four gap genes; inset shows hetereogeneous Bicoid concentration inside a single nucleus (airy-scan image with experimental design described in Ref. \cite{Munshi_2024}). B) The normalized concentration of the four gap gene proteins (\textit{hunchback}, \textit{giant}, \textit{Krüppel}, \textit{knirps} in red, yellow, blue, green respectively) in nc 14 as a function of embryonal position x, scaled to embryo length L, exemplifying the developing body segments \cite{Petkova}. C) \textit{hb} RNA expression in nc 13 as a function of embryonal position \cite{Littleetal}, when \textit{hb} expression is mostly regulated through the proximal hbP2 enhancer region. D) Sketch of real the hbP2 region \cite{Park_2019}, containing ca. 6 binding sites, and our simplification with equidistant binding sites of equivalent binding strength.
\label{fig0}}
\end{figure}

\subsection{Clustering and Information in Early Fly Embryo Development}
The Bicoid morphogen gradient is 
the most easily experimentally accessible of the maternal signals 
that are required for correct body-part segmentation in the fly embryo (Fig. \ref{fig0}A) \cite{driever1988gradient, Gregor2007a, Liu}. The information provided in these maternal signals is transferred downstream via the gap genes, whose expression begins consistently after twelve rounds of nuclear divisions, in nuclear cycle (nc) 13.
In nuclear cycle 14, these gap gene patterns become fully refined (Fig. \ref{fig0}B) and provide enough information for nuclei to develop into distinguishable cells \cite{Dubuis, Petkova}. This information is then transferred towards larger group of genes, including the segmentation genes, which determine cell fates and scale with embryo size \cite{Dubuis, Nikolic}. Although gap proteins act as transcription factors with mutually activating and repressive functions \cite{Jaeger_2011} 
that contribute towards refining their pattern,
in nc 13 the gap gene \textit{hunchback} (\textit{hb}) can be considered as being regulated predominantly by Bicoid.

Bicoid has disordered or low-complexity domains, which are associated with condensation \cite{KatoMcKnight}. These condensation phenomena are most relevant for high concentrations, and transcription factors are often expressed at low (nM) concentrations:  Concretely, Bicoid concentrations correspond to approximately 10000 molecules at the anterior (head) and 0-10 molecules at the posterior (tail) of the embryo \cite{Gregor2007a, dostatniBicoid1, Mir}. Yet, also at these lower concentrations, clustering can occur through affinity to different, highly-expressed and clustering proteins, or in the form of pre-wetted, local clusters only around binding sites \cite{morin2022sequence,Munshi_2024,Mir_2018, Dufourt_2018}. 
Indeed,  hubs or clusters involving Bicoid have been observed experimentally \cite{ Mir,Mir_2018, Dufourt_2018, Munshi_2024}.

The expression of \textit{hb} in nc 13 (Fig. \ref{fig0}C) is driven predominantly 
by the P2 (proximal) promoter or enhancer region (hbP2), 
with approximately six core binding sites (Fig. \ref{fig0}D). This region has frequently served as a test region for gene expression \cite{Tkacik_2008, Ling, Park_2019, Eck2020, Desponds2020, neri2025amendable}. 
Recently, mRNA expression statistics for all gap genes, including from this region, have been measured on the level of single polymerases \cite{ChenLevoZoller}. 
Here, we consider a simplified version of this enhancer as a bottleneck variable for information transfer in the context of transcription factor clustering.  We study a single regulatory region with approximately equi-distant binding sites with equal binding energies, and take the occupation of this binding site sensor as the variable that determines cell fates downstream. We investigate only the local environment around this binding site sensor along the DNA, ignoring the three-dimensional microenvironment, potential competition effects between sensors or refining effects from shadow enhancers \cite{Perry2010, bothma2015enhancer,Park_2019}.

The occupation $C$ of the binding site sensor corresponds to the nucleus's measurement of the signal $s$, and can be mathematically expressed as the averaged fraction of occupied sites during a time interval $\tau$: $C = \frac{1}{\tau}\int_{0}^{\tau} c(t) dt$, with $c$ denoting the occupation at time $t$. We consider downstream gene expression to be approximately proportional to $C$, consistent with simple models of gene regulation, in which $C$ can determine the number of polymerases recruited to the promoter. 
This approach neglects promoter noise, which is acceptable for weak promoters \cite{weinert2014scaling}, and is consistent with equilibrium thermodynamic models of gene regulation \cite{Bintu}. Recent work points towards a necessity for non-equilibrium models in gene regulation \cite{estrada2016information,scholes2017combinatorial,WongGunawardena,zoller2022eukaryotic,shelansky2024single,tkavciktenwolde,Zoller2025}. While including polymerases in ways that match experimental promoter statistics for this region \cite{ChenLevoZoller,Zoller2025} presents an exciting direction for future work, the absence of non-equilibrium effects in our model affects our estimates for the time interval $\tau$ during which binding site sensors can measure and respond.

Estimating a realistic value for measurement time $\tau$ is difficult, since our model does not include all processes in the nucleus. We therefore consider two limiting measurement times: a comparatively short time, corresponding to one or a few measurements that determine gene expression for the rest of the nuclear cycle, and a long measurement time, which represents the maximum possible realistic timescale at this stage of development, corresponding to the entire lengthscale of nc 13 with generous parameter estimates. Non-equilibrium effects, such as de-coupling between occupation and transcription, chromatin accessibility or transcription factor rebinding would shorten this timescale estimate. Therefore, we expect that the realistic measurement time interval $\tau$ lies in between our two limiting timescales. 

\subsection{Information Flow Calculations}
To assess how much information about the external signal of concentration $s$ the regulatory region (or sensor) can capture, we use the mutual information
\begin{equation} \label{eq:info}
I(C;s) = \int ds \int dC P(C,s) \log_2 \frac{P(C,s)}{P(C)P(s)}\textrm{,}
\end{equation}
 which depends on probability distribution $P( C|s)$, describing the nucleus's measurement of $s$.
 
We use the Bicoid expression profile \cite{Gregor2007a} (Fig. \ref{fig0}A) to obtain the prior distribution of signalling molecules $P(s)$, by taking $P(s|x)$ Gaussian and $P(x)$ uniform. For stochastic models of binding at a binding site, we can obtain $P(C|s)$ from simulations (section III). For sufficiently long time intervals $\tau$, longer than the correlation time,  the probability distribution $P(C|s)$ approaches a Gaussian distribution, i.e.,
\begin{equation} \label{eq:gauss}
P(C|s) = \frac{1}{\sqrt{2 \pi \sigma_{C}(s)^2}} \exp \Big[ -\frac{(C(s)-  \bar{C}(s)  )^2}{2 \sigma_{C}(s)^2} \Big]\textrm{,}
\end{equation}
where $\sigma_{C}(s)^2$ is the variance and $\bar{C}$ the mean occupation that occurs during the interval $\tau$. For such long $\tau$, $\bar{C}$ corresponds to the mean occupation in steady state, $ \bar{C} :=  \langle C \rangle =\frac{1}{\tau} \int_{0}^{\tau} \langle c \rangle dt =  \langle c \rangle$, where $\langle \ldots \rangle$ denotes the ensemble average. 
 We can calculate $\sigma_{C}(s)^2$ and $\bar{C}$ for a specific stochastic process, and use this Gaussian approximation to assess information transfer through a specific binding site. 

The information provided by the signal could be different from the information that the embryo needs to develop correctly. For example, a signal may provide more information than needed in specific instances. In the fly embryo, correct development requires that nuclei along the embryonal axis can develop into distinct fates, determined by their position \cite{Dubuis,tkavcik2021many}. We thus use the information that the regulatory region can convey about nuclear position $x$, $I(C;x)$ \cite{Dubuis, Baueretal}, as a variable of relevance to the embryo. We calculate $I(C;x)$ using Eq. \ref{eq:info} with $P(C|x) = \int P(C|s) P(s|x) ds$, consistent with the information bottleneck framework \cite{tpb00} and the simplified problem we address here, where only Bicoid provides information about cell fates to the enhancer. We can therefore assess precision in the transfer of molecular signals through gene regulation with both the information captured about the signal, $I(C;s)$ as well as the information of relevance, $I(C,x)$, allowing for a more nuanced assessment than possible with  signal-to-noise ratios.

\section{Simple model for gene regulation: weak cooperativity is optimal  \label{sec:Hill}}
\subsection{Hill function models: Obtaining probability distributions}
\begin{figure*}
 \centerline{\includegraphics[width=0.85\textwidth, keepaspectratio]{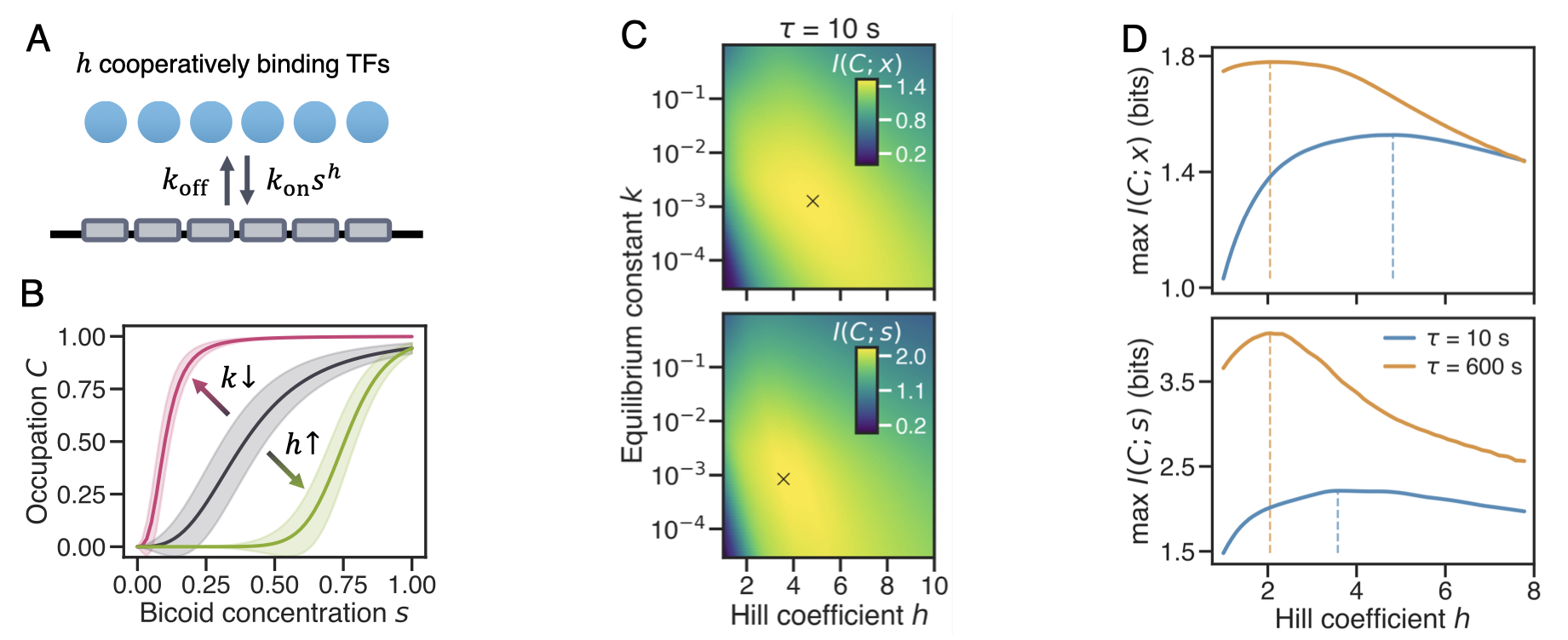}}
\caption{Information transfer through a sensor with cooperative binding sites. A) Sketch of cooperative binding of transcription factors (TFs) to six binding sites. B) Mean occupation according to the Hill function for different parameters $h$ and $k$. C) The information $I(C;x)$ (and $I(C;s)$) as a function of $h$ and $k$ shows a well-defined maximum (cross) in parameter space. We use $k_\textrm{off} = 1/$s. D) The maximum possible information $I(C;x)$ and $I(C;s)$ over all values of $k$ for different $h$ (maxima indicated by vertical lines). For the longer measurement time $\tau=600$ s (orange), the maximal information value per $h$ is higher than for $\tau=10$ s (blue), and maximal values are reached for lower $h$.   \label{fig:1} }
\end{figure*}

In  models for gene regulation, the mean occupation of binding sites $\bar{C}$ for cooperative binding is frequently described by a Hill function \cite{Hill, deJong,weiss1997hill, bialekbook, TkacikWalczak, Marzen, Bauer_2022},
\begin{equation} \label{eq:hill}
\bar{C}   = \frac{s^h}{k+ s^h} \textrm{.}
\end{equation}
where $s$ is the normalized signal concentration, $k$ the dissociation equilibrium constant and $h$ the Hill coefficient corresponding to the number of binding sites or cooperatively binding molecules (Fig. \ref{fig:1}A,B).
Hill functions  originate from a heuristic deterministic description that compares the average number of free and bound binding sites over long time intervals, and accurately describe gene expression outputs in population-averaged experiments. 
Using Hill function models for gene expression in single cells is more complicated due to the lack of an entirely realistic mechanism, which we address with a stochastic description in section \ref{sec:mechanism}. Nevertheless, Hill functions capture gene expression phenomenologically: sigmoidal activation profiles for gene regulation that can be matched to Hill functions are frequently observed \cite{Bintu, scholes2017combinatorial, zoller2022eukaryotic}; Hill functions with $h=1$ correspond to MWC models, also canonically used for gene regulation \cite{Marzen}; and finally, Hill functions provide an interesting limiting case for a stochastic model as the steepest possible binding site occupation for a regulatory region with $h$ binding sites in single cells \cite{Martinez2024hill}. Therefore, we explore Hill function models here to guide expectations. 

First, we design a stochastic process for which the mean occupation can be represented with a  Hill function, which will allow us calculate the variance in occupation.
We assume that binding only occurs when $h$ molecules bind at the same time to $h$ binding sites (Fig. \ref{fig:1}A). The probability for the binding site to be unoccupied at time $t$, $P(c=0,t)$, can be described by the master equation for the two states of being occupied or not occupied, $c=1$ and $c=0$, 
$\frac{ d P(c=0,t) }{dt } 
%&
= k_{\textrm {off}} - (k_{\textrm {off}}+k_{\textrm {on}}s^h) P(c=0,t)\textrm{,}
%\end{align}
$
where $k_{\textrm {on}}$ and $k_{\textrm{off}}$ are the rate constants for binding and unbinding, and we used $P(c=1,t) = 1-P(c=0,t)$
The mean occupation in steady state for this process corresponds indeed to a Hill function (Eq. \ref{eq:hill}) with equilibrium constant  $k=k_{\textrm{off}}/k_{\textrm {on}}$, if the time-averaged normalized mean occupation $\bar{C}$ is approximately the ensemble-averaged, steady-state mean, $\bar{C} = \langle c \rangle$.

We assume that the probability with which a binding site sensor  is occupied during a time interval at specific signal concentration $s$ can be approximated with a Gaussian distribution (Eq. \ref{eq:gauss}) with this occupation mean and 
the variance  $\sigma_{C}(s)^2 = \langle (C - \bar{C} )^2 \rangle $. An estimate of this variance can be obtained from the observation that the regulatory region is either occupied or unoccupied during each measurement; in time $\tau$, there will be of order $N \approx \tau/ \tau_c$ independent measurements, where the correlation time 
\begin{equation} \label{eq:tauc}
\tau_c = \frac{1}{k_{\textrm {off}} + k_{\textrm {on}} s^h}
\end{equation} can be obtained from the master equation. Therefore, we expect a binomial variance of order $\sigma_{C}(s)^2 \approx \frac{1}{N} \bar{C}(1 - \bar{C})$, where in steady state $\bar{C} (1 -\bar{C})$ corresponds to the instantaneous variance \cite{Kaizu, tenWolde, bialekbook}. We follow the calculation in Ref. \cite{Kaizu} to obtain the exact result (Appendix \ref{sec:appA}), 
\begin{align} \label{eq:var3}
\sigma_{\bar{C}}(s)^2  =\frac{2 \tau_c }{\tau}  \bar{C} (1- \bar{C}) \textrm{,}
\end{align}
 in agreement with Refs. \cite{Kaizu, tenWolde, bialekbook}. This differs from the ad-hoc expectation by a factor of two; this factor is required when the concentration $s$ is inferred from the occupation during the time interval $\tau$, whereas inferring the concentration from the discrete number of (independent) binding events allows for a lower variance \cite{EndresWingreen}.

The variance in occupation (Eq. \ref{eq:var3}) increases with the correlation time of the occupation, which in turn increases with Hill coefficient $h$ (Eq. \ref{eq:tauc}). Therefore, one might expect that cooperativity in the form of $h>1$ reduces the information transfer, consistent with Refs \cite{BialekSetayeshgar2,SkogeWingreen, SkogeII}.

With these expressions, we can calculate  how precisely a binding site sensor can measure signal concentration $s$, provided the sensor is determined by parameters $h$ and $k$ and has access to a time-averaged occupation during a particular interval $\tau$. We calculate the distribution for the occupation $P(C|s)$  using Eqs. \ref{eq:hill}, \ref{eq:var3} and \ref{eq:gauss}. We investigate two measurement times representative of a short and long measurement $\tau$. The long measurement corresponds approximately to the duration of nc 13 (10 min), provided that transcription factor residence times of order $\approx 1$s  can set an effective timescale for  $k_{\textrm {off}}\sim 1/\text{s}$,
consistent with measurements from Ref. \cite{Mir}. The short measurement is still longer than the correlation time, to ensure that the Gaussian approximation is valid; we chose a time of approximately $\tau=10$ s.

\subsection{The information  about the signal and relevant function increases for weak clustering.}

Both the information about the signal $I(C;s)$ as well as the information of relevance to the embryo, $I(C;x)$ show a clear maximum at a specific parameter combination of $h$ and $k$ per measurement time (Fig. \ref{fig:1}C, D, Appendix \ref{sec:appA}). This optimum occurs at finite $h$ for all measurement times we studied, and for the longer of our two limiting $\tau$, values close to this optimum are achieved by a range of cooperativities $h$ (exact optimum at $h \approx 2.1$, Fig. \ref{fig:1}D top). The presence of this optimum at finite $h$ is in agreement with earlier work that also incorporated noise in RNA production \cite{tkavcik2009optimizing}. 

Both $I(C;x)$ and $I(C;s)$ increase with the measurement time $\tau$, both when optimizing over $h$ and $k$  (Appendix \ref{sec:appA}), and when just optimizing over $k$ as a function of cooperativity $h$ (Fig. \ref{fig:1}D). This is expected, since the variance decreases with measurement time. For long $\tau$, the information of relevance $I(C;x)$ saturates at approximately $1.8$ bits (Appendix \ref{sec:appA}), as the binding site cannot extract more information about cell fates than the signal, i.e. the Bicoid concentration,  provides (additional signals would allow for more information \cite{Dubuis}). In this long measurement time limit and even for the unrealistic limit of $\tau \to \infty$, the optimal cooperativity saturates at $h\approx 1.38$ (Appendix \ref{sec:appA}).

Even if we compare the cooperative measurement to one where $h$ binding sites are treated independently, cooperative measurements with $h>1$ are optimal (not shown).
We find therefore that for realistic timescales, weak cooperativity is optimal independent of whether information about signal or cell fates are considered.

From an information-theoretic perspective, it makes sense that steep thresholds corresponding to high $h$ are optimal for shorter measurement times,  associated with high noise or variance in $P(C|s)$:  binary, threshold-like measurements are known to provide the most information when noise is high \cite{Baueretal, Mattingly2018, witteveen2025}. The optimum occurs at finite cooperativity since the variance is nevertheless low enough for $\tau = 10$s to allow for a steep but not strict thresholded mean. For longer $\tau$  and more precise measurements, we find that weak cooperativity is optimal or does not reduce information up until $h\approx 3$ (Fig. 2D orange). This may seem surprising, since one might expect a linear activation mean to become optimal as noise increases. However, all activation means we study here, even for $h=1$, imply a non-linear sigmoidal $\bar{C}=s^h/(k+s^h)$, and such non-linear functions make it difficult to think about information-theoretic optima intuitively. In addition, for considering $I(C;x)$, we should note that the Bicoid gradient itself varies non-linearly as a function of $x$;  rate-distortion theory predicts that an optimal map $\bar{C}(s)$ that compresses a signal to convey information about $x$ should be a non-linear function of the signal if $x$ is conveyed by the signal with a non-linear map $\bar{s}(x)$ (concretely $\bar{C}(s) =\bar{x}(s)$ \cite{CoverThomas}). Since optimizing $I(C;x)$ is additionally a convex problem, different non-linear maps can achieve the optimum.

If the longer measurement time of 10 minutes is biologically realistic for the fly embryo, an optimally adjusted binding site sensor would have approximately 2-3 cooperatively binding molecules. Interestingly, this does indeed correspond to the number of strongly binding Bcd molecules in the hbP2 enhancer \cite{Park_2019}. We do not expect an exact match, as our estimates for timescales are approximate, but are encouraged to find a reasonable estimate for $h$. We conclude that our heuristic Hill-function model suggests that some cooperativity is optimal, consistent with information-theoretical expectations considering the functional form of the signal. Yet, 
the Hill function model suffers from the significant shortcoming that we are forced to consider longer measurement times and cannot capture the binding and unbinding of individual molecules. To check the role of clustering in more detail, we next investigate 
a more realistic model that proposes a regulatory mechanism based on clustering transcription factors.

\section{Mechanistic model for clustering transcription factors \label{sec:mechanism}}

\begin{figure*}
  \centerline{\includegraphics[width=\textwidth]{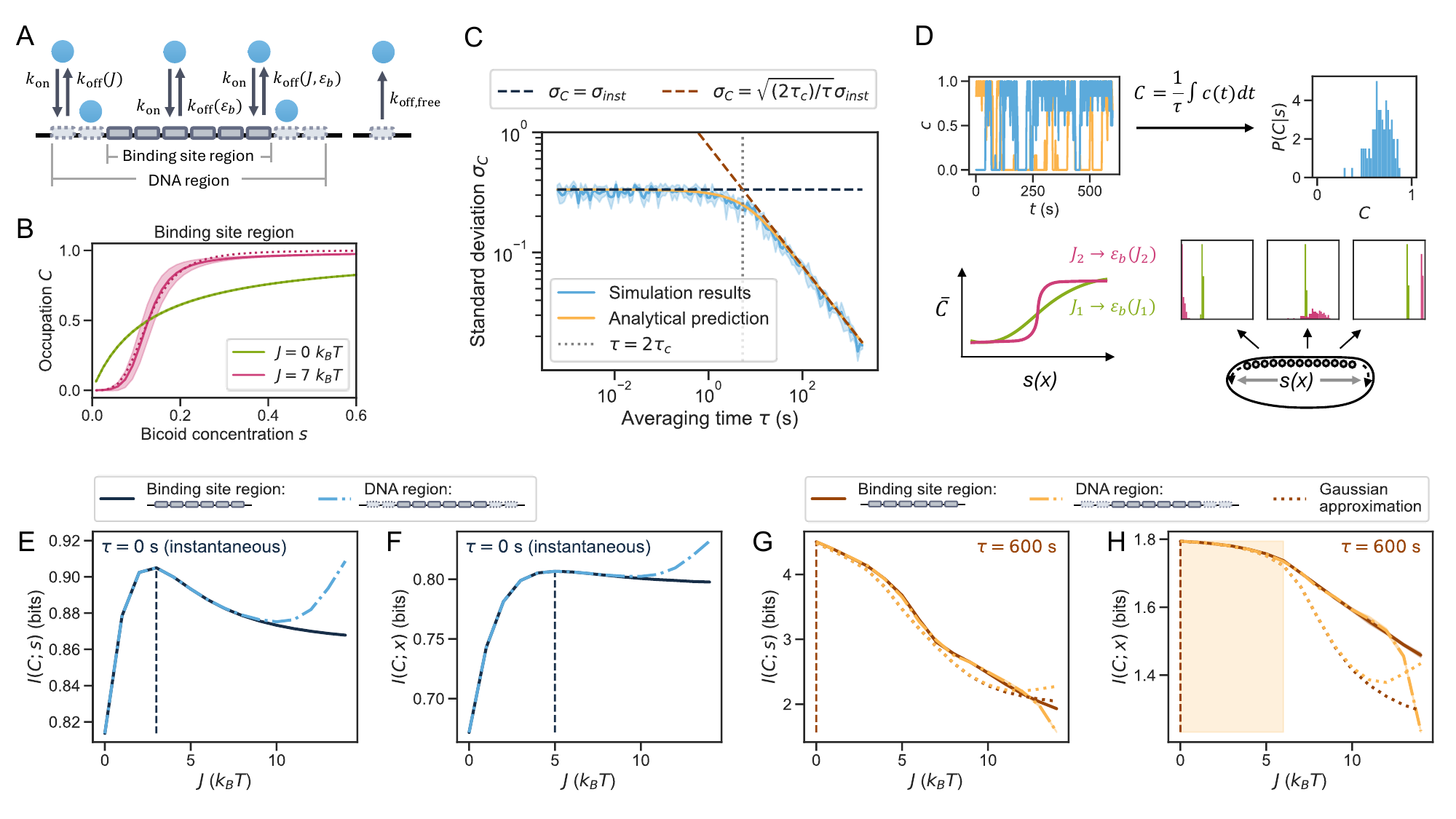}}
\caption{Information transfer through a binding site or DNA region with individual clustering signal molecules. A) Sketch of model binding sites (solid) and surrounding sites on DNA (dashed; DNA region), with on- and off-rates for transcription factors. B) Steady-state occupation $\bar{C}$ of the binding site region (no surrounding sites) for $J=0$ and $\varepsilon_b\approx 14.1$, and  $J=7$ and $\varepsilon_b\approx8.2$ as a function of signal concentration, $s$, with the standard deviation (shaded region) for measurement time $\tau=10$ min, and best-fit Hill function (dotted line), with $h\approx 1$ and $h\approx 4$, respectively. 
C) Standard deviation $\sigma_{C}$ from
simulations of the binding site region matches the analytical solution from the master equation (which can be approximated by Eq. 7 as shown in Appendix VI C, Fig 7A), decreasing from its instantaneous value to  a  $1/\sqrt{\tau}$-scaling for longer $\tau$. 
Parameter values are $\mu =-14.0$, $J=3.5$ and $\varepsilon_b\approx 11.1$, with error bars over 5 independent repeats of the calculation, each based on 20 simulations.
%\textcolor{blue}{(Parameter values $\mu =-14.0$, $J=3.5$ and $\varepsilon_b\approx 11.1$; error bars over 5 independent repeats of the calculation, each based on 20 simulations.)}
 D) Sketch of the simulation procedure: for all $\mu$ corresponding to signal concentrations $s(x)$ along the embryonal axis, we perform Gillespie simulations and average occupation over time $\tau$ to obtain $C$. For each $s$, we estimate  $P(C|s)$ based on histograms of $C$ from 100 repeats of the simulation. We calculate $I(C;s)$ and $I(C;x)$ from this distribution for each $J$, using a modified parameter $\varepsilon_b$ to retain half maximum occupation at $x\sim 0.47$. E+F) Mutual information in an instantaneous measurement: 
Clustering improves both $I(C;s)$  and $I(C;x)$, with  a maxima for  clustering strengths  $J\approx 3$ and $J\approx 5$, respectively. For high $J$, the occupation of additional sites along the DNA can further increase the information. G+H)  Mutual information in a long measurement ($\tau=600$ s): Clustering decreases $I(C;s)$. The information $I(C;x)$ also decreases with clustering, but with minor information loss (5\% of the maximum) up until intermediate clustering strengths of $J=6$ 
(shaded area). \label{fig:2} }
\end{figure*}

\subsection{Adapted Ising model for binding sites}

% Goal: introduce the model
To find a more realistic description of the occupation of a binding site sensor that incorporates binding and unbinding of individual transcription factors, we adapt an Ising model for a small, one-dimensional region of the DNA around the binding sites (Fig. \ref{fig:2}A).  We fix the concentration of the signal transcription factors with chemical potential $\mu$, parameterize the binding energy for a binding site with $\varepsilon_b$, and describe clustering between neighbouring transcription factors with clustering strength $J$. 
We calculate the energy of a specific configuration of transcription factors using
\begin{equation}
E(\{c_1, \ldots c_N\})=  - J \sum_{i=1}^{N-1} c_{i} c_{i+1} - \mu \sum_{i=1}^{N} c_i - \varepsilon_b \sum_{i=1}^{N'} c_i \textrm{,}
\end{equation}
where $c_i= \{0,1\}$ denotes occupation at site $i$ and $N$ ($N'$) denotes the number of (binding) sites in the system, and where we refer to $J$, $\mu$ and $\varepsilon_b$ in units of $k_BT$. The energy of  configurations determines the rates for individual events that can occur, which we implement consistent with detailed balance \cite{Glauber}.
We simulate two types of events, corresponding to transcription factors appearing from and leaving to a bath (the nucleus), with rates $k_{\textrm{on}}$ and $k_{\textrm{off}}$. Since arrival statistics should not depend on local effects or the identity of the site (binding site or DNA site), we use the same $k_{\textrm{on}}$ for all sites. This rate $k_{\textrm{on}} = e^{\beta \mu} k_{\textrm{off,free}}$ (with $\beta=1$) then only depends on signal concentration $s$ (through $\mu$) and on a global timescale in the system (Appendix \ref{sec:appB}), which is set by $k_{\textrm{off,free}}$, the rate with which transcription factors leave an isolated or free DNA site (Fig. 3A). We incorporate the local configuration, including site identity and clustering effects with a neighbouring transcription factor, into the rate $k_{\textrm{off}}$ (Appendix \ref{sec:appC}).

A core difference between cooperativity and clustering is that a clustered hub of proteins can extend beyond the binding site region. Thus, we study two types of binding site sensors: first, a binding site sensor with $N'=N=6$ binding sites (Fig. \ref{fig:2}A, binding site region), and second, an extended binding site sensor with $N'=6$ binding sites surrounded by two sites on each side that represent the surrounding DNA region, such that $N=10$ (Fig. \ref{fig:2}A, DNA region). 
We will see that an interesting effect of the DNA region sensor is that steeper thresholds are achievable for strong clustering strengths $J$, which can be beneficial for short measurement times. 

We obtain 
probability distributions for the normalized time-averaged occupation $C$ of the binding site sensor (referred to as `occupation' from now on) 
with Gillespie simulations \cite{gillespie1976,gillespie1977} (Appendix \ref{sec:appD}). 
To calculate the information, we again focus on limiting measurement durations that represent a short and long measurement time. We use an instantaneous measurement and a long timescale, corresponding again to an estimate of the maximal possible timescale in nc 13, ca 10 min, which we estimate using diffusion constants and by generously matching lengthscales in our simulation to those in the fly embryo (Appendix \ref{sec:appB}, \cite{dostatniBicoid1,Fernandes_2022,Gregor_2007b,Porcher_2010}).

However, to guide expectations, we first explore the mean $\bar{C}$ and variance $\sigma_{C}^2$ of the occupation $C$ 
as a function of input parameters. We refer to Appendix \ref{sec:appC} for the stochastic master equation and detailed calculations.

\subsection{Clustering causes steeper occupation mean and increase in occupation variance.}
The mean occupation $\bar{C}(s)$ increases more steeply 
with signal concentration $s$ 
for higher clustering strength $J$ (Fig. \ref{fig:2}B). Clustering also shifts the signal concentration at which the mean occupation of the binding site sensor reaches its half-maximal occupation; varying the binding energy can compensate for this shift (Fig. \ref{fig:2}B). 
A Hill function can approximate the mean occupation, justifying our approach in the first section. 

The variance around the mean occupation $\bar{C}$ in steady state can be obtained using the correlation function (see Appendix \ref{sec:appC}, Ref. \cite{SkogeWingreen, ZhangBerry}). We find
\begin{align}\label{eq:Variance}
\sigma_{C}^2 \approx 2 \sigma_{\textrm{inst}}^2  \frac{e^{-\tau/\tau_c} + \frac{\tau}{\tau_c} -1 }{(\tau/\tau_c)^2}\textrm{,}
\end{align}
where  $\sigma_{\textrm{inst}}^2$ is the instantaneous variance in steady state, which describes the fluctuations in (instantaneous, not time-averaged over $\tau$) occupation around the steady state mean at stationarity, $\sigma_{\textrm{inst}}^2 = \langle (c - \langle c \rangle)^2 \rangle$. This expression reduces to $ \sigma_{C}^2 (\tau) \approx \sigma_{\textrm{inst}}^2 \text{ for } \tau \ll 2 \tau_c$ and to $ \sigma_{C}^2 (\tau) \approx \frac{2 \tau_c}{\tau} \sigma_{\textrm{inst}} \text{ for } \tau \gg 2 \tau_c$,  
as expected \cite{Kaizu}.
We validate this approximation by showing excellent agreement of the variance from Eq. \ref{eq:Variance} with a semi-analytical expression using the master equation (Appendix \ref{sec:appC} Fig. \ref{fig:ts}A,B), which matches the stochastic simulations (Fig. \ref{fig:2}C).
Clustering increases the correlation time of a measurement, which in turn increases the variance (Appendix  \ref{sec:appC}, \cite{SkogeWingreen}). 

We thus conclude that the strength with which transcription factors cluster, $J$, affects both the mean and variance of the occupation at the binding site sensor. While stronger clustering leads to increased variance, the concomitant effect on the mean occupation may foil the conclusion that clustering always decreases information, just as in section II.

\subsection{In the fly embryo, weak clustering can lead to an increase in mutual information about cell-fates.}

Before
presenting how clustering affects information transfer in the fly embryo in this model, we explain one modification that we make to reduce the parameter space which we sample. In our model, increasing the clustering strength $J$ shifts the
mean occupation towards anterior parts of the embryo. This behaviour is consistent with the idea that clustering increases the local concentration of transcription factors \cite{Hnisz_2017, TrojanowskiRippe}. Yet, as a consequence, many combinations of $J$ and the binding energy $\varepsilon_b$ lead to sensors that capture barely any information. Such sensors are not biologically realistic: If a particular clustering strength is set by properties of transcription factor proteins, and if the information transfer from Bicoid is important, the embryo would likely have faced evolutionary pressures that ensure that the binding energy is adjusted through mutations in binding sequences, to make information transfer feasible.  
Therefore, as we vary $J$ in the following section, we choose to vary $\varepsilon_b$ at the same time, setting $\varepsilon_b$ such that the binding site sensor can produce half maximal expression at $x\approx 0.47$, as expected for \textit{hb} expression in the fly embryo (Fig. 1C, Appendix \ref{sec:appC} Fig. \ref{fig:ts}E). This reduces the parameter combinations we study. While we do not prove that this choice is information-theoretically optimal, we see in section IV that it is close to the information bottleneck bound. For each $J$, we obtain mutual informations by calculating all $P(C|s)$ for values of chemical potential $\mu(x)$ that yield a particular signal (Bicoid) concentration $s$ (Fig. \ref{fig:2}D).

For instantaneous  measurements, weak clustering increases information transfer: both $I(C;s)$ and $I(C;x)$ increase with $J$ with  local maxima at $J\approx 3$ and $J\approx 5$, respectively (Fig. \ref{fig:2}E,F). This is consistent with our results from Hill-function models, and the intuition that steeper, more thresholded measurements are optimal when the measurement is noisy. 
This robust maximum at finite clustering strengths, consistent across models, is noteworthy. It occurs at finite clustering strength, not at a sharp threshold or infinite clustering, since the instantaneous variance is finite and we are not in large-noise limit where strictly binary measurements are optimal (Appendix \ref{sec:appC}). 

For the DNA region sensor that includes a  larger region around the binding site, stronger clustering of $J\approx 10$ improves  information transfer again. This increase in information is related to the mean occupation: the effective Hill coefficient for the binding site region reaches $h=6$ from below for high $J$ (Appendix \ref{sec:appC} Fig.\ref{fig:ts}E), 
since the Hill coefficient is bounded by the number of binding sites \cite{EstradaWong,Martinez2024hill}; in contrast, the steepness for the DNA region can increase further, as clusters extend beyond just the binding sites.
The information will eventually saturate at higher $J$, also for longer DNA sensors, beyond the valid regime of our simulations: For $J\gg15$, the binding energy $\varepsilon_b$ required to keep half-maximal occupation at the center of the embryo becomes unrealistically small. 
%Although a more accurate calculation would require extension of our model with multiple sensors in the canonical ensemble 
We used  an artificial sensor with periodic boundary conditions to estimate that information will saturate at  $I(C;s) \approx 0.93$ and $I(C;x) \approx 0.85$ in our model.

The fact that clustering can lead to occupation of surrounding DNA sites could therefore be an avenue by which clustering can increase information transfer in the embryo (see also Refs. \cite{rippe2022,Pontius1993}). However, for signals expressed at low concentrations, such strong clustering is likely also associated with a source of error that our model does not capture: for small numbers of signal molecules and multiple binding site regions, large clusters around one region may deplete signal molecules from  other binding site regions, and therefore contribute to a more noisy measurement in realistic settings. Since we do not take into account these depletion effects, we cannot evaluate this effect here.

For long measurement times, clustering reduces the 
information about the signal (Fig. \ref{fig:2}G). This information loss occurs because of a combination of two effects: since measurements are more precise (low variance) for long $\tau$, the steeper mean activation of occupation that increasing $J$ facilitates is less beneficial. In addition, clustering additionally increases the variance, as the increasing correlation time implies that less  independent measurements can be performed during interval $\tau$, consistent with earlier work \cite{SkogeWingreen}. 
An important contrast is the information of functional relevance to the embryo: $I(C;x)$ exhibits a plateau for weak clustering strengths and does not decrease significantly (within five percent) for clustering up to $J\approx6$ (Fig. \ref{fig:2}H). This information is barely affected for weak $J$, since the effect from the increasing variance can be partially compensated by the fact that for conveying relevant information, a steeper non-linear mean can present an optimal map if the map between signal and relevant variable is also non-linear. Therefore, this result validates and refines our findings from section \ref{sec:Hill}, and 
 shows that also for a mechanistic model, clustering corresponding to Hill coefficients up until $h\approx 3$  is not detrimental to relevant signal processing. 

For long $\tau$, we can verify the validity of the Gaussian approximation, which we used in section \ref{sec:Hill}, in comparison to simulations: it works well until medium clustering strengths, $J\approx6$, above which $P(C|s)$ becomes too strongly peaked at empty or full occupation (Fig. \ref{fig:2}G,H). We also observe that 
the difference between the extended DNA sensor and binding site region is marginal for the long measurement time, as the information is not limited by the steepness of the mean occupation. 

Finally, we emphasize again that we expect realistic timescales for the embryo to lie between the instantaneous and the long measurement time. Therefore, our results show that weak clustering is consistent with optimizing the functionally relevant information transfer.

\section{The information bottleneck bound is achievable with realistic binding site sensors \label{sec:ib}}
\begin{figure*}
 \centerline{\includegraphics[width=.95\textwidth]{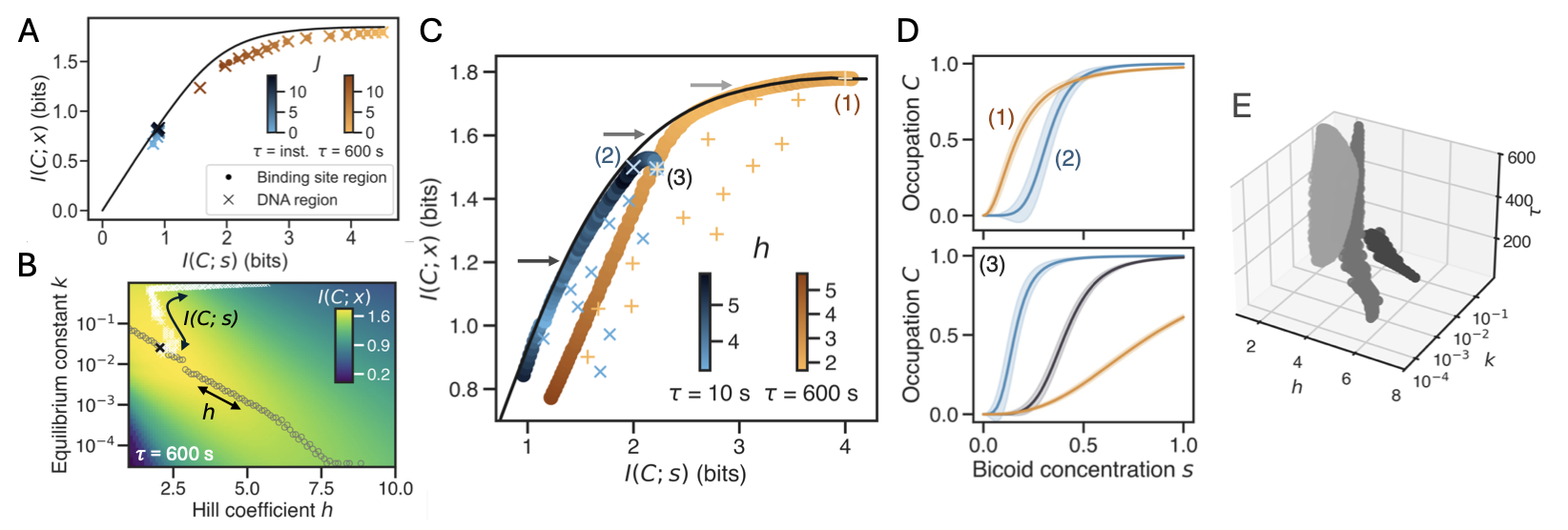}}
\caption{Binding site sensors that incorporate cooperativity and clustering can theoretically achieve the information bottleneck optimization goal.  A) Sensors from the mechanistic model for instantaneous (blue) and long measurement times (orange) for changing clustering strength $J$ (color shade) on the information plane, with optimal IB bound from data in black. B) Different binding site sensors from the Hill function model, parameterized by $h$ and $k$, are optimal for different constraints: grey circles and white crosses mark paths with a (variable) constraint on $h$ and a (variable) constraint on $I(C;s)$ (IB goal), respectively.
 C) Information plane with IB bound in black and a selection of exemplary binding site sensors (blue crosses and orange plusses for $\tau=10$ s and $\tau=600$ s, respectively) show that randomly chosen binding sites are well below the bound. The sensors that optimize $I(C;x)$ for a given $I(C;s)$ in our parameter ranges of $h$ and $k$ are shown with larger circles; color shade indicates the value of $h$. D) Selected binding site sensors (1,2) from panel C representing the sensors along the IB bound (top), and sensors that represent point (3) in panel C (bottom, for $\tau=10$ s in blue, $60$ s in grey and $600$ s in orange). E) Optimal values of $h$, $k$ and $\tau$ for binding site sensors fitted to $P(C|s)$ from IB optimization, for three values on the bound in Fig.4C (grey arrows) show that a variety of binding site sensors are consistent with the IB bound. 
\label{fig:ib} }
\end{figure*}

In the previous sections, we discussed $I(C;x)$, as well as the information that the binding site can capture about the signal, $I(C;s)$ as variables that quantify 
maximal information flow. 
In this section, 
we study information maximization with an optimization goal that involves a trade-off between these information quantities, as described by the information bottleneck (IB) problem \cite{tpb00}.

The information transfer from a molecular signal via a binding site region to downstream gene expression can be viewed as 
an IB problem \cite{tpb00}:  the binding site sensor $C$ presents a bottleneck between a signal $s$ and a variable of importance to the embryo, the position along the embryonal axis $x$, describing possible cell fates downstream \cite{Baueretal, BauerBialek}. The distribution $P(C|s)$ can be optimized such 
that $I(C;x)$ is maximal 
for a specific value of binding site capacity $I(C;s)$, constrained by Lagrangian multiplier $\lambda$:
\begin{equation}\label{eq:IB}
\underset{P(C|s)}{\textrm{max}} \quad I(C;x) - \lambda I(C;s) \textrm{.}
\end{equation}  
For each value of $\lambda$ and therefore for each value of $I(C;s)$ we can find the optimal  
sensor and the information $I(C;x)$ it can convey (Appendix \ref{sec:appE}), without making assumptions about the sensing mechanism. This optimal 
IB bound 
is obtained directly from the 
distribution that maps the relevant variable to the signal, $P(s|x)$. 
The optimal bound of maximal $I(C;x)$ for each $I(C;s)$  divides the information plane (Fig. \ref{fig:ib} A,C) into an allowed region on or below the optimal IB bound and a forbidden region above. 

The quantity constrained in the IB problem is $I(C;s)$, which is determined by 
the physical parameters describing the binding site sensor, including the number of binding sites, clustering strength $J$ or binding energy $\varepsilon_b$, as well as diffusive timescales or the local environment of the binding site. In absence of knowing which of these parameters are constrained, a constraint on $I(C;s)$ can serve as a reasonable general constraint. 
The IB bound has recently been explored in the context of molecular sensing, and has predicted several features of enhancer networks that appear in the fly embryo as optimal, such as a need for multiple enhancers or a combination of binding sites for different signals \cite{Baueretal}. 
Now, we investigate how clustering of transcription factors affects the ability of binding site regions to operate close to this bound.

We compare realistic binding site sensors  to this bound by showing pairs of $I(C;s)$ and $I(C;x)$ from binding site sensors with specific parameter combinations and models on the information plane. We begin with binding site sensors from the mechanistic simulations, for both measurement times and varying $J$ (Fig. \ref{fig:ib}A). Strikingly, all binding site sensors are close to the information bound, for all values of $J$. 
Sensors with longer measurement times $\tau$ measure more precisely and occupy regions higher in the information plane than sensors with instantaneous measurements, which also cover a smaller region of phase space (orange vs blue in Fig. \ref{fig:ib}A). 

In general, we see sensors close to the IB bound as consistent with IB optimality, while sensors that are not optimal fall far below the bound; this is because we do not know a priori which $I(C;s)$ is physically possible and therefore what specific value point or range of values in the information plane are reasonable. Some binding site sensors are within $10\%$ of the maximally achievable $I(C;x)$ (light orange); these sensors involve weakly clustering transcription factors, up to clustering strengths of $J\approx 3$ (Fig. \ref{fig:ib}A). More specifically, the elbow of the IB plot marks a particular `sweet spot' in optimization. We find that for Bicoid, an abstract IB-optimal sensor can operate within 20\% of the bound of 1.8 bits when sensing with a capacity of 1.8 bits. We further find that sensors with finite clustering, $J\approx 5$, are closest to this elbow of the plot. This suggests that transcription factor clustering is beneficial especially from an IB perspective, since clustering helps the sensor optimize a trade-off by measuring as much relevant information as possible with the noisiest possible measurement. 

 It may be surprising that all binding site sensors from the mechanistic model for all values of clustering $J$ are close to the IB bound. Our choice of constraining $\varepsilon_b$ for each $J$ to fix the half-maximal occupation to the middle of the embryo thus implicitly presents an effective optimization consistent with the information bottleneck.  While variation of $\varepsilon_b$ may have allowed us to push even closer to the bound, it is interesting that our biologically motivated choice satisfies the bound well,  as it suggests that this joint tuning of clustering strength and binding site energy can provide a smooth, potentially evolutionarily accessible path towards optimizing the relevant information.

To explore more binding site sensors, we return to the Hill-function models from section \ref{sec:Hill}. First, we visualize the effect of changing the constraint from fixed $h$ to fixed $I(C;s)$ (Fig. \ref{fig:ib}B). The optimal sensor parameters differ 
between constraints, intersecting at the optimum. 
Along the IB constraint $h$ initially increases for decreasing $I(C;s)$,
 followed by a decrease until the equilibrium constant $k$ is so large that the occupation of an individual binding site sensor barely reaches 50\% for the highest $s$ (Appendix \ref{sec:appE}). This suggests that forcing sensors for a single measurement time to occupy all parts of the IB plot is not sensible. Therefore, we do not consider binding site sensors with maximal occupations below 0.5.  

Binding site sensors with varying $h$ and $k$ can fall on a wide region below the IB bound (symbols in Fig. \ref{fig:ib}C), for both $\tau$. Sensors that measure longer achieve more information, especially for more optimized sensors (orange crosses extending to higher $I(C;x)$). To investigate whether some binding site sensors match the IB bound, we numerically select the binding site sensor that  maximizes $I(C;x)$ for each $I(C;s)$ (as in \ref{fig:ib}B), and show the corresponding information pairs in the information plane (blue and orange lines, with color shade indicating $h$, Fig. \ref{fig:ib}C). The optimal sensors are on the IB bound for different regions of phase space: the sensors that measure for longer $\tau$ 
are unable to 
achieve the IB bound for low $I(C;s)$, due to our constraint on a sensor's mean occupation. Nevertheless, the sensors for short $\tau$ achieve the bound for low $I(C;s)$ and are in general steeper than sensors for long $\tau$ close to the bound (Fig \ref{fig:ib}D top).

Different sensors can yield the same $I(C;s)$ and $I(C;x)$ and therefore end up at the same region of the information plane (Fig. \ref{fig:ib}D bottom). 
Indeed,
 multiple parameter combinations $h$, $k$ and $\tau$ yield sensors with the same $I(C;s)$ and $I(C;x)$ along the bound (Fig. \ref{fig:ib}E).
 This shows that multiple implementations of sensors can be consistent with information-theoretic optimization goals \cite{opt+var}; nevertheless, randomly selected sensors are not optimal.

\section{Discussion and Conclusion \label{sec:dis}}
\subsection{Considering functionally relevant information resolves the paradox of clustering in info-max principles.}
We developed two models that allow us to calculate the precision with which binding site sensors measure transcription factor signals, and the information about cell fates that the binding site sensor can capture from the signal. %
Our first model based on Hill-functions relies on a series of approximations, and does not describe the binding of individual molecules. To go beyond this, we developed a second more realistic model that also allows us to study clustering around a binding site region.
We chose the occupation of a binding site region, for which a decrease in the signal-to-noise ratio due to clustering had previously been predicted, as the variable that needs to convey information downstream. 
In both models, the presence of cooperativity or clustering can increase the relevant information
that the binding site region can convey for short measurements, as steep enhancer activations are optimal for noisy sensing. Even for the longest measurement duration that could possibly be realistic according to our estimates, clustering does not significantly reduce this relevant information.  

Our results are not sensitive to the details of signal transfer at the binding site: we verified that weak clustering is even more beneficial for a binding site region that induces \textit{hb} expression not proportionally to its average occupation, but only when it is fully occupied \cite{Desponds2020} (Appendix \ref{all-or-nothing} Fig. 8).
We therefore find that when drawing conclusions about the optimality of information transfer in a biological system, it is important to consider the functionally relevant information, rather than an estimate about the signal, which led to an apparent conflict of info-maximization principles based on previous estimates.   

We also showed that binding site regions with clustering transcription factors  are able to
 achieve information optimality as predicted by the information bottleneck bound  with clustering pivoting the sensor towards a particularly beneficial sensing tradeoff.
Here, we showed that a variety of different binding site sensors are IB optimal. The good performance of a single sensor may be surprising, but this likely reflects our maximum timescale estimate being too generous, or our rate constants not being realistic. This suggests that it will be interesting to identify biologically realistic constraints on parameters, which could include molecule numbers or energy constraints \cite{sokolowski25,Tjalma, Tjalma2, nicoletti2024tuning}, and investigate their impact on IB optimality.

\subsection{Connection to current experiments.}
Our result that clustering can be beneficial for information transfer
connects with current work on the importance of clustering around binding site regions in the context of transcription: clustering can buffer signal fluctuations, which improves accuracy for a constant signal that fluctuates \cite{Klosin, Deviri}. Here, we work towards understanding  how clustering  affects the processing of signals whose concentration conveys information. 
 Our findings provide an information-theoretical understanding of the usefulness of clusters that stretch between and around binding sites \cite{rippe2022,Pontius1993}, and of the importance of weak interactions in the context of transcription \cite{gao2018evolution}. 
While current work for larger transcriptional hubs has discussed a Goldilock's principle of a narrow regime of optimal clustering strength \cite{fallacaro2025, chong2022tuning}, with our model we find that multiple clustering strengths can be close to optimal information transfer if binding energies are adjusted. 

In order to allow for a more direct assessment of enhancer and promoter regions with respect to information-optimality, three improvements to our work should be considered. First, our estimates of timescales suffer from diffulties with matching to new data for residence times and occupation statistics of transcription factors \cite{Mir, Munshi_2024, fallacaro2025}. A model improvement that would make this easier involves extending the size of our system to a larger 3D geometry with different binding site sensors. This geometry could then allow for transcription factors diffusion, and could capture competition effects between these sensors: If clustering occurs at multiple regulatory regions at the same time, 
transcription factors could be depleted \cite{Nimwegen} and these 
competition effects between these binding sites regions will reduce the accuracy of information transfer.  Conversely, some regulatory regions regulate expression combinatorially \cite{furlonglevine_18}. 
 
Second, the promoter statistics should be included more explicitly in our model. Recently, mRNA production of the gap genes has been measured on the level of individual polymerases \cite{ChenLevoZoller}. Models that reproduce these polymerase statistics suggest that an out-of-equilibrium drive is required in a model that includes polymerases \cite{Zoller2025}. An exciting direction is to add polymerases with non-equilibrium rates, to obtain promoter regions that match experimental statistics, and investigate their optimality following our pipeline. We expect that adding polymerases would significantly reduce information capacities in our system, but that it would provide clearer insights into how tightly squeezed biological systems are for achieving optimal information transfer. 

Third, a more complete picture of information transfer would require incorporating additional proteins and combinatorial inputs, and potentially their dynamics. Multiple maternal signals provide information to multiple outputs, including the four gap genes, assisted by pioneer factors \cite{limErk,harden2023transcriptional}. This combinatorial regulation can allow for modular optimal solutions that may fulfil biological functions and depend on steeper activation profiles, which clustering could help facilitate. In addition, polymerases and pioneer factors can cluster themselves \cite{cho2018mediator, Mir}, and it has recently been suggested that variables other than binding site occupation assist in carrying relevant information, such as cluster shape or size \cite{mittag2022conceptual, Mir_2018, Munshi_2024}. Finally, the temporally changing dynamics of these proteins could present an exciting avenue for considering information in the full trajectory \cite{Reinhardt, Moor}, including temporal smoothening of output profiles through diffusion \cite{sokolowski2015optimizing}, or the opening of enhancer regions due to pioneer activity \cite{schulz2019mechanisms,birnie2023precisely,harden2023transcriptional}.

These improvements would allow us to assess whether real regulatory regions with clustering transcription factors optimally processes information. For the single binding site region that we study here, a test for the connection between clustered occupation and downstream processing would involve measurements of transcription factor occupation at this region (ideally labelled \cite{beliveau2015single, gizzi2019microscopy}) in combination with concomitant output (RNA and protein). Experimental possibilities to tune the clustering strengths between molecules could further allow us to understand the landscape of optimal parameters that are still consistent with healthy development \cite{opt+var}.

\textit{Acknowledgements.} 
We are grateful to Ned Wingreen, Eric Wieschaus and William Bialek for many helpful discussions of questions, models and calculations relevant to this project. We thank Thomas Gregor and Mikhail Tikhonov for providing data for \textit{hb} mRNA for Fig. 1C, and Rahul Munshi for providing an image of Bcd in Fig. 1A.  We thank Riccardo Rao, Linda Dierikx, Max Metlitski, Caroline Holmes, Michal Levo, Thomas Gregor and his lab, Trudi Sch\"upbach and members of the Bauer group for comments and discussions. We acknowledge funding from the NWO Vidi Talent programme (NWO/VI.Vidi.223.169, MB), the NWO Science-XL grant OCENW.XL21.XL21.115 (AK and MB), and a TUDelft Start-up grant (JZ). A part of this work was performed at the Aspen Center for Physics, which is supported by National Science Foundation grant PHY-2210452. 

\newpage 

\section{Appendix}
\subsection{Effective Hill function model: Variance and numerics \label{sec:appA}}
For the Hill function model, we assume that binding only occurs when $h$ molecules bind at the same time to $h$ binding sites. The chemical master equation for the two states of occupation at the binding site, $c=0$ and $c=1$, reads
\begin{align} \label{eq:ME1}
\frac{ d P(c=0,t) }{dt } %&= k_{\textrm{off}} (1-P(c=0,t)) - k_{\textrm {on}} s^h P(c=0,t) \nonumber \\
&= k_{\textrm {off}} - (k_{\textrm {off}}+k_{\textrm {on}}s^h) P(c=0,t)\textrm{,}
\end{align}
where $k_{\textrm {on}}$ and $k_{\textrm{off}}$ are the rate constants for binding and unbinding. We assume that the signal concentration  $s$  is not affected by binding, since the number of binding sites is typically much less than the number of signal molecules in the nucleus. 
We can solve the master equation, 
\begin{align}
P(c=0,&t)= \frac{k_{\textrm{off}}}{k_{\textrm{off}} + k_{\textrm{on}}s^h} + \nonumber\\&\left(P(c=0,t=0) - \frac{k_{\textrm {off}}}{k_{\textrm{off}} + k_{\textrm{on}}s^h}\right) e^{-(k_{\textrm{off}} + k_{\textrm{on}}s^h)t}\textrm{,}\nonumber
\end{align} 
to find the mean at steady state. This mean indeed corresponds to the Hill function, 
\begin{equation} 
\bar{C} =\langle c \rangle  = \frac{s^h}{k+ s^h} \textrm{,}
\end{equation}
as in the main text.
Although the effective ergodic, long $\tau$ approximation that we need for the first equality ($\bar{C} = \langle c \rangle$) may not hold for single nuclei and shorter timescales, we proceed with this approach, as Hill-functions are established in gene regulation, and present an interesting and accurate limit for the true stochastic process, as described in the main text.

To calculate the variance, we consider the temporal correlation function at stationarity: $\kappa (\delta t) =  \langle ( C (0) -  \bar{C}) ( C( \delta t) - \bar{C}) \rangle$. This correlation function depends only on the time difference $\delta t$ \cite{Gardiner}. 
This allows us to write for the variance
\begin{align}\label{eq:var1_1}
\sigma_{C}(s)^2 & = \langle (C -  \bar{C} )^2 \rangle\textrm{,} \nonumber \\& = \frac{1}{(\tau)^2} \int_{0}^{\tau} 
  \int_{0}^{\tau} 
\langle ( C (t_1) - \bar{C}) ( C(t_2) -  \bar{C}) \rangle  d t_1 d t_2\textrm{,} \nonumber \\ &
= \frac{1}{(\tau)^2} \int_{0}^{\tau} \int_{-t_1}^{\tau - t_1} \kappa( t) d t dt_1 \textrm{,}
\end{align}
following the calculation in Ref. \cite{Kaizu}. We swap the order of integration in Eq. \ref{eq:var1_1} and exploit symmetry of $C(t)$ to perform the integration over $dt_1$:
 \begin{align} \label{eq:var2_1}
\langle (C - \bar{C})^2 \rangle &= \frac{2}{\tau^2} \int_{0}^{\tau}  dt \kappa(t) (\tau -t) \textrm{.}
\end{align} 
This can be approximated further if the correlation function $\kappa(t)$ decays on a timescale shorter than the measurement time $\tau_c \ll \tau$,
 \begin{align} \label{eq:var3_0}
\sigma_{C}(s)^2 = &\langle (C - \bar{C})^2 \rangle \nonumber  \approx \frac{2}{\tau} \int_{0}^{\tau} dt \kappa(t) \textrm{.}
\end{align}
 We can obtain $\kappa(t)$  from the solution from the master equation, Eq. \ref{eq:ME1},
\begin{equation}
\kappa(t) =\bar{C} (1- \bar{C})e^{ -  t/\tau_c}=k_{\rm off} k_{\rm on } s^h \tau_c^2 e^{ -  t/\tau_c} 
\end{equation}
 where $\tau_c$ is the correlation time from Eq. \ref{eq:tauc}, and $\bar{C} (1-  \bar{C})$ is the instantaneous variance as in the main text. Using $k=\frac{k_{\textrm {off}} }{k_{\rm on} }$, the variance can be expressed as
\begin{align} \label{eq:var3_1}
\sigma_{C}(s)^2 & = \frac{2}{\tau k_{\rm on}}\frac{ k s^h}{(k + s^h)^3} \textrm{,}\nonumber\\
& = \frac{2}{\tau k_{\rm off}}\frac{ k^2 s^h}{(k + s^h)^3} \textrm{,}\\
& = \frac{2 \tau_c }{\tau} \bar{C} (1- \bar{C}) \textrm{,}\nonumber
\end{align}
for large $\tau$. 

\begin{figure}[ht!]
%\centerline{\includegraphics[width=0.45\textwidth]{Figures/Supplement/Figure2C_Supp_kon10constant.pdf}}
\centerline{\includegraphics[width=0.5\textwidth]{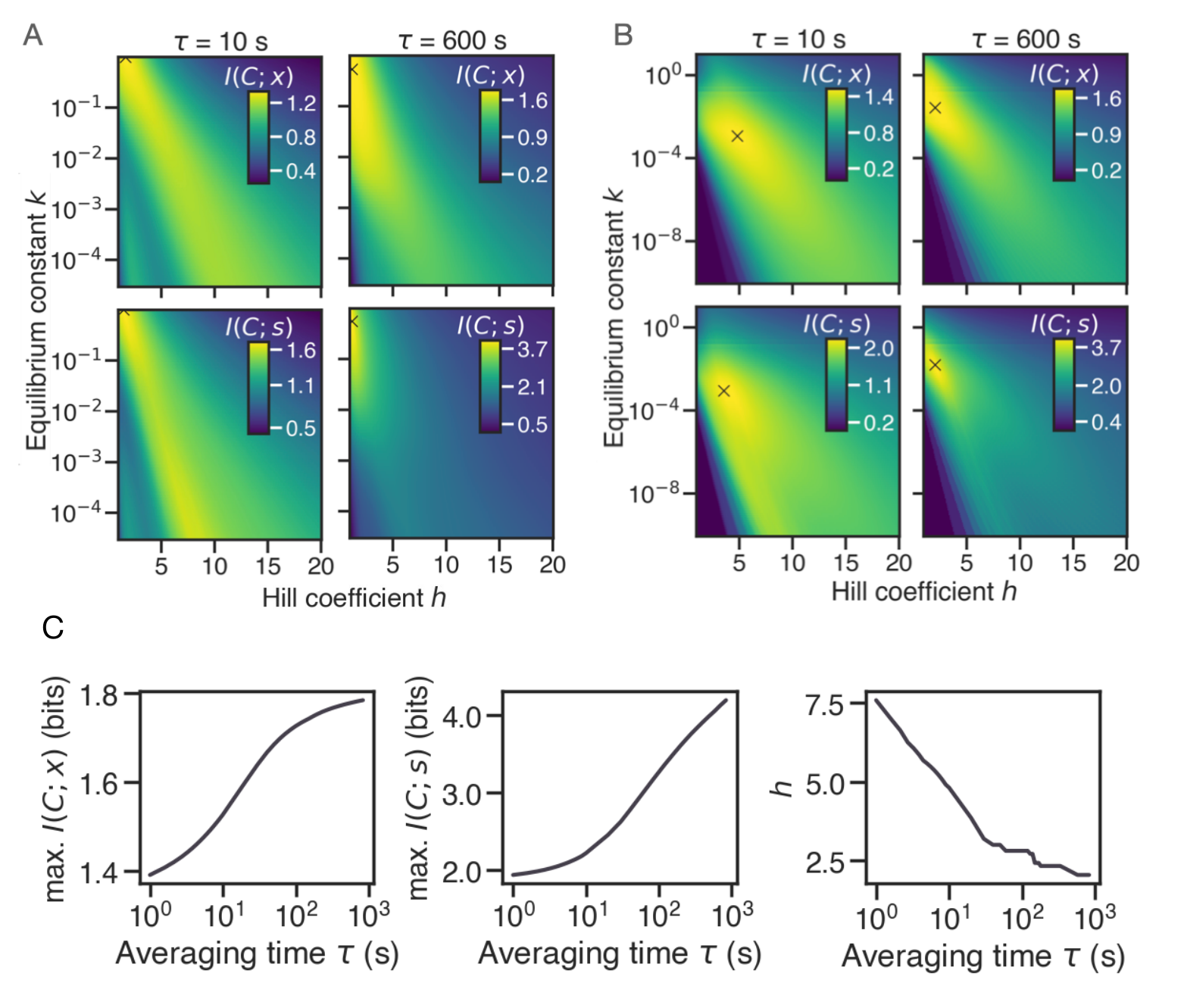}}
\caption{$I(C;x)$ and $I(C;s)$ as a function of $h$ and $k$ for two representative measurement times. A) $I(C;x)$ and $I(C;s)$ for $k_{\rm on} = 10$ s$^{-1}$ nM$^{-h}$ constant and varying $k_{\rm off}$; the maxima (cross) are always at the boundaries of parameterization space. B) Longer measurements with $\tau = 600$ s show a similar behaviour to $\tau=10$ s, with a single optimum (cross). The left panels are replotted from the main with an increased range of $k$ values. C left and middle) The optimized information $I(C;x)$ and $I(C;s)$ over all $k$ and $h$ as a function of measurement time $\tau$, with $I(C;x)$ reaching its maximal bound $I(s,x)$ at approximately $\tau\sim 1000s$. C right) The value of $h$ at the optimum of $I(C;x)$ decreases with measurement time $\tau$. \label{fig:suppa2}}
\end{figure}

We use Eq. \ref{eq:var3_1} to calculate the variance for our binding site sensors depending on parameters $k =k_{\rm off}/k_{\rm on}$, $h$ and $\tau$. The system depends on three free parameters, which can equivalently be chosen as $\tau k_{\rm on}$, $k_{\rm off}$ and $h$, or as $\tau k_{\rm off}$, $k_{\rm on}$ and $h$ (see Eq. \ref{eq:var3_1}). We opt for the second choice, as the first options errs towards a fallacy where the information can always increase as $k_{\rm off}$ increases (Fig. \ref{fig:suppa2}A with the maximum at the boundary of our range for $k$): in our system, the information to the binding site is provided only through the binding, and therefore a fast unbinding rate (and instantaneous disappearing of the molecule in our diffusive-free system) allows for more independent measurements. We therefore make the practical choice to keep $k_{\rm off}=1$ fixed and vary $k_{\rm on}$.

For comparison, Fig. \ref{fig:suppa2}B shows the information provided through a binding site with varying $h$ and $k$ (for constant $k_\textrm{off}$, as in the main text) for both measurement times side-by-side. The maximum information, indicated by a cross, occurs for finite $h$ and $k$.

%\begin{figure}[h]
%\centerline{\includegraphics[width=0.45\textwidth]{Figures/Supplement/Figure2C_Supp_extendedkrange.pdf}}
%\caption{$I(C;x)$ and $I(C;s)$ for a longer measurement time of $\tau = 600$ show a similar behaviour to those for $\tau=10$, with a single optimum (cross). The left panels are replotted from the main with an increased range of $k$ values.\label{fig:suppB}}
%\end{figure}

\begin{figure}
\centerline{\includegraphics[width=0.45\textwidth]{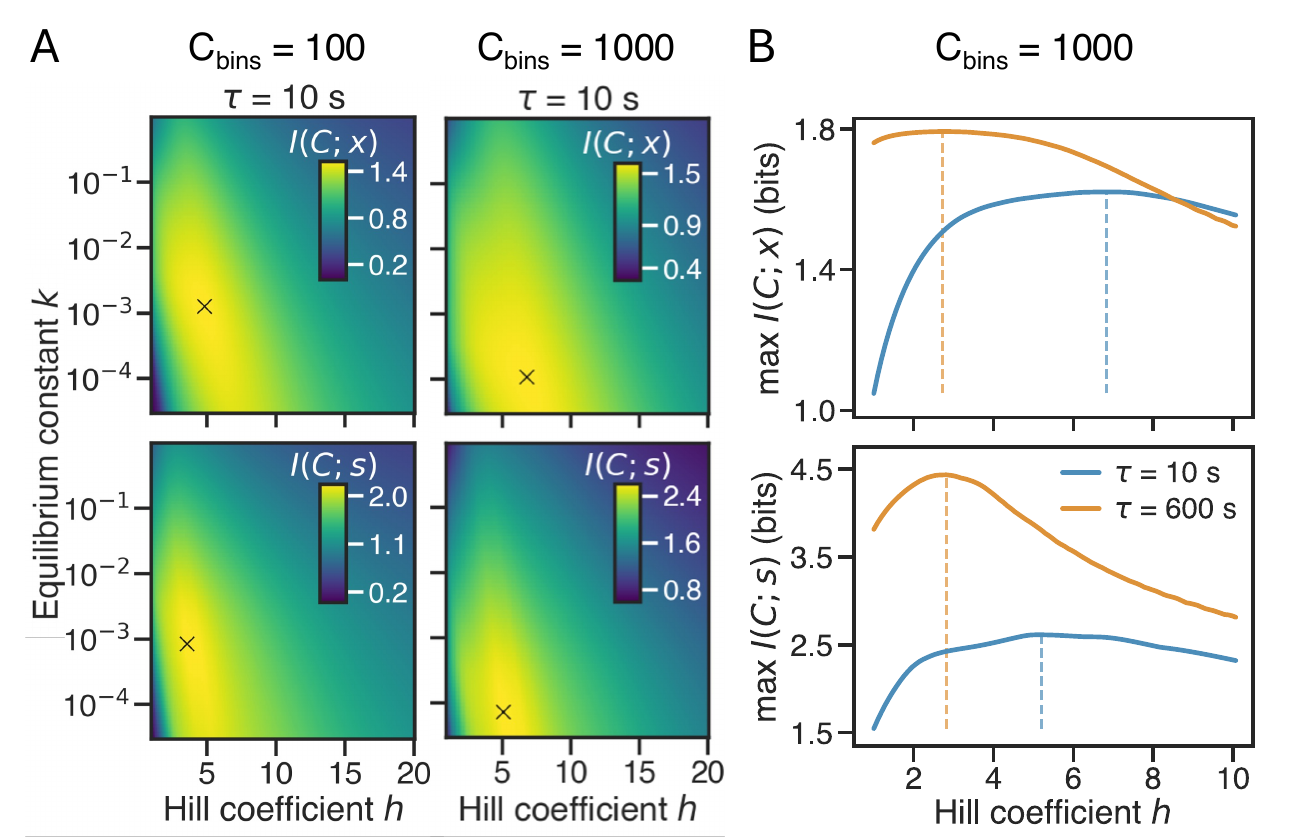}}
\caption{$I(C;x)$ and $I(C;s)$ using a changed number of bins for the numerical calculation of $P(C|s)$. A) The information $I(C;x)$ and $I(C;s)$ as a function of $h$ and $k$  for $\tau=10$ s shows maximum for binding site parameters (cross) at finite $h$, independent of numerical details. These details can affect the exact parameterization of the optimum, especially the dissociation equilibrium constant $k$. B) The maximum possible information $I(C;x)$ and $I(C;s)$ over all values of $k$ for different $h$. Finite $h$ is optimal for information transfer and the $h$ at the optimal mutual information $I(C;x)$  decreases with $\tau$, as in the main manuscript (Fig \ref{fig:1}D).\label{fig:suppC}}
\end{figure}
We calculate probability distributions using the Gaussian assumption numerically, with 100 bins in the variable $C$ in the main text. To ensure that the finite number of bins does not affect our result, we recalculated the information as a function of binding site parameters $h$ and $k$ for a different numerical parameterization, with 1000 discrete bins for the variable $C$.  We observe minor shifts for the parameters $h$ and $k$ that make up the maximum (Fig. \ref{fig:suppC}), but the results do not change otherwise; therefore, we decided to use 100 bins for the main text.  

We use 160 positions between 0.1 and 0.9 embryo length (for details see \ref{sec:appD} for information about x-position bins).

\subsection{Time and length scales for the mechanistic model \label{sec:appB}}
The length scale of the model is set by the shortest distance between the centers of two Bicoid binding sites in the \textit{hb} enhancer: 12 bp or approximately $4$ nm \cite{Fernandes_2022}. The shortest distance between the centers of two Bicoid binding sites in the \textit{hb} enhancer is 12 bp or approximately $4$ nm \cite{Fernandes_2022}.
We thus
use a grid cell of 4 nm.  With this grid size, the
concentrations used in our simulations are all below $10^{-12}$ molecules/grid cell, and thus satisfy the
condition that the occupation at each binding site is much less than 1. 

We assume that on the surrounding sites in the DNA region the transcription factors do not interact with DNA. Therefore, the
off-rate in the absence of binding and clustering depends only on the timescale in which a transcription factor diffuses away. To simplify the calculations, we rescale time such that $k_\textrm{off,free}$ is equal to 1 per time unit. One time unit is then equal to the mean time it takes a transcription factor to diffuse through the grid cell. We roughly estimate this timescale using the equation for the mean squared displacement (MSD) of a diffusing particle:
\begin{equation}
\textrm{MSD} = 2KDt \to t = d_x^2/(2KD)
\end{equation}
where $D$ is the diffusion constant, $d_x$ is the grid cell size and $K$ is the number of dimensions in
which the transcription factor can leave the grid, with $K = 3-1$ for a one dimensional grid. Values for
the diffusion constant of Bicoid in literature range form 0.3$\mu m^2/s$ \cite{Gregor_2007b} to 7.7$\mu m^2/s$ \cite{Porcher_2010}. We use the more likely and recent value of $ D \approx 7 \mu m^2/s$ as found by \cite{dostatniBicoid1}. With these values we obtain a time unit of t=$1/k_\textrm{off,free} \approx 5.7 \times 10^{-7}$ seconds, or $k_\textrm{off,free} = 1.8\times 10^6/s$.
With typical binding energies that we use for weak clustering, these timescales typically correspond to $k_\textrm{off,free}\sim 1/s$, which is consistent with the timescale estimate we use in the Hill function model. Residence times of order 1s are consistent with recent experiments for Bicoid \cite{Mir}, as well as previous experiments in the fly embryo but also in other systems \cite{Porcher_2010, Hansen2017, chen2014, kim2021single, fountas2024better}. We note that we assume that the estimate for our longer measurement is likely longer than what is realistically possible for enhancers: this is partially because we chose generous estimates for diffusion and grid size, and because the enhancer is likely not actively measuring for the entire duration of the cycle.

\subsection{Semi-analytical calculation of standard deviation  for the mechanistic model\label{sec:appC}}
In the absence of clustering, $N$ states with different discrete occupations $n$ are enough to describe a system with $N$ sites. For a system with $N=6$ binding sites, the probability of the state with $n$ transcription factors bound at a timepoint $t$, $P_n (t)$ is given by
\begin{align}
    \dfrac{dP_n(t)}{dt} &= [(N-(n-1))k_{\textrm{on}}]P_{n-1}(t) + [(n+1)k_{\textrm{off}}]P_{n+1}(t) \nonumber \\ & - [(N-n)k_{on} + n k_{\textrm{off}}]P_n(t),\quad \textrm{for} \quad 0<n<N, \nonumber \\
 \dfrac{dP_0(t)}{dt} &=k_{\textrm{off}} P_{1}(t) - N  k_{\textrm{on}} P_{0}(t), \nonumber \\
\dfrac{dP_N(t)}{dt} &=k_{\textrm{on}} P_{N-1}(t) - N  k_{\textrm{off}} P_{N}(t),
\end{align}
where $P_0$ and $P_N$ describe probabilities for empty and full states. The variable $n$ is related to $c$ in the main manuscript by a normalization constant, i.e. $c=n/N$. Probabilities need to satisfy the constraint $\sum_{n=0}^{N} P_n (t) = 1$. 
The expression for the energy of a configuration in the main manuscript can be used to determine the energies of a particular state and thereby
 determine the rates $k_{\textrm{on}}$ and $k_{\textrm{off}}$, where $k_{\textrm{on}}$ only depends on signal concentration $s$ and through it on $\mu$, and where $k_{\textrm{off}}$ depends on the site identity and the local neighbourhood, 
\begin{equation}
  k_{\textrm{on}} = e^{\beta \mu} k_{\textrm{off,free}} \text{ and } k_{\textrm{off}} = e^{-\beta (\varepsilon_b+Jn)} k_{\textrm{off,free}}.
\end{equation}
In the limit of a single binding site, we recover Eq. \ref{eq:ME1} with $h=1$ and, using $\mu = k_B T \ln s$, the steady state mean corresponding to the Hill function (Eq. \ref{eq:hill} in the main manuscript, i.e. $\frac{s}{s+ e^{-\beta \varepsilon_b}}$).

To incorporate clustering, we need to take into account all possible states of occupation in what could be seen as a graph-theoretical approach \cite{WongGunawardena}: different states with the same occupation are connected with different rates. Taking into account all states requires a $2^{N} \times 2^{N}$ transition matrix with elements $M_{ij}$, where $i$ and $j$ run over all $2^N$ different states, described by their occupation for each binding site $\{n_1,\ldots,n_N\}$, where each $n_i\in\{0,1\}$. For example, at a particular time, a binding site sensor might be in state $\{1,1,1,0,0,0\}$. We denote the occupation of each such state by $\tilde{c}_i =\sum_{k=1}^{N} (\{n_k\})/N$, so that $\tilde{c}$ is also a $2^N$ vector.

The probability to occupy state $i$, $\tilde{P_i}$, can be obtained from the solution of the master equation, via 
\begin{equation} \label{eq:ME2} \tilde{P_i}(t) = \sum_{j=1}^{2^N} \Big ( e^{M t} \Big)_{ij} \tilde{P_j}(0) \textrm{,}\end{equation}
where $\tilde{P_j}(0)$ denotes the initial probability of state $j$, which we take to be the steady state probability. The transition matrix $M$ has one zero eigenvalue, corresponding to steady state. 
Since all transition matrices at thermodynamic equilibrium can be transformed into a symmetric transition matrix $\tilde{W}$ with the same eigenvalues  \cite{ZhangBerry}, we know that this matrix is diagonalisable and will continue below with the representation using orthonormal eigenvectors, following references \cite{SkogeWingreen, ZhangBerry}. 

With the eigenvalues of $M$, $\lambda_k$, we can write $ \dfrac{1}{\tau}\int_0^\tau e^{M t}dt = \dfrac{1}{\tau} U A U^{-1} $, where $A$ is a diagonal matrix, the $k$ diagonal elements of which correspond to the integrals $ \int_0^\tau e^{\lambda_{k} t}dt$ over all eigenvalues of $M$ and where $U$ is the matrix of eigenvectors of $M$. Since the first eigenvalue is zero, corresponding to the steady state eigenvalue, we obtain $A_{11} = \tau$ and  $A_{ii} = \frac{1}{\lambda_i }( e^{ \lambda_i \tau}-1)$ for $i>1$. This allows us to find the expectation value of the time-averaged occupation $C$,   
\begin{figure*}[ht]
 \centerline{\includegraphics[width=\textwidth]{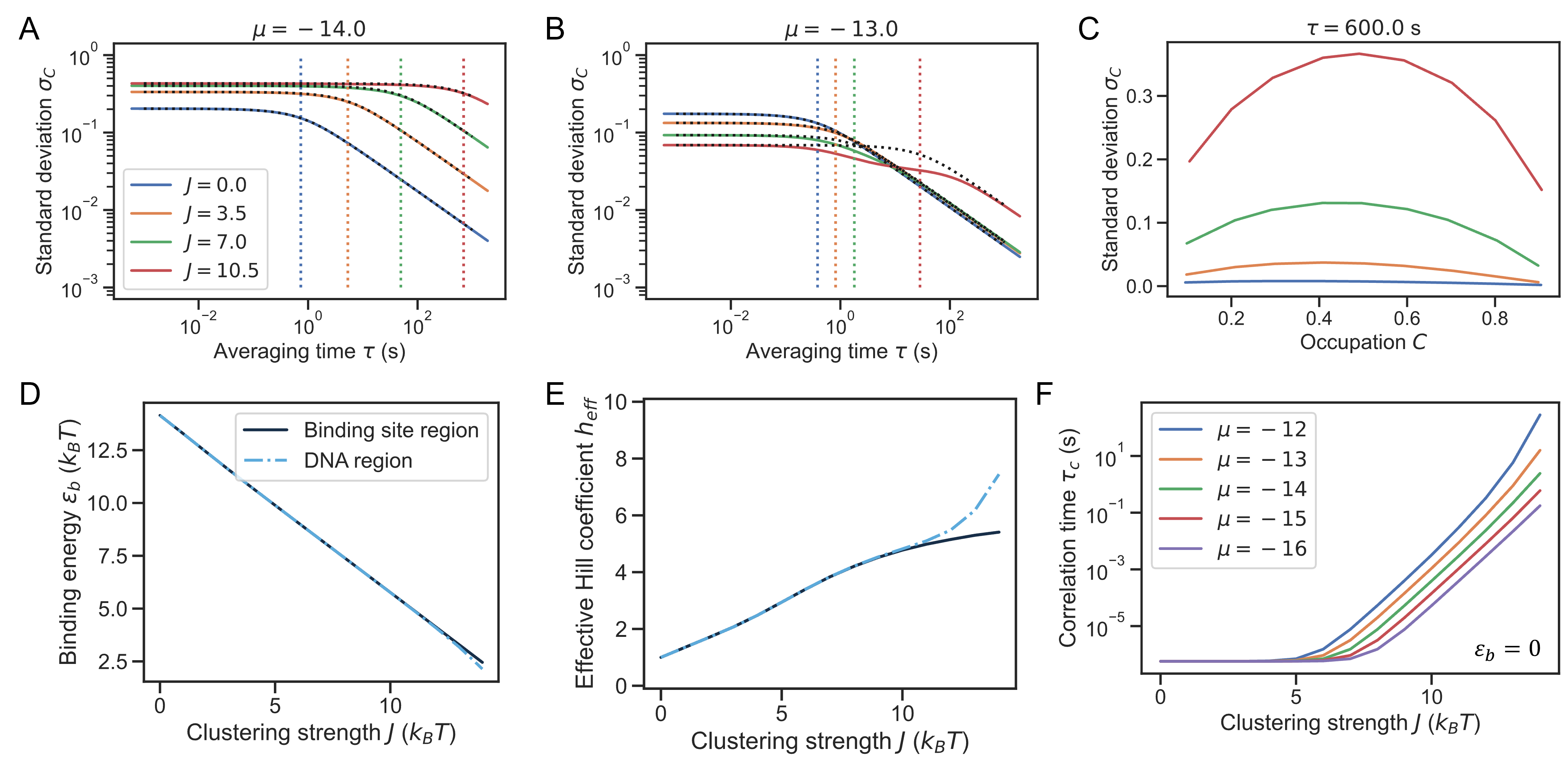}}
\caption{Mean and standard deviation of the mechanistic model from the master equation and analytic approximations. A+B) Comparison of standard deviation from diagonalization of Eq. \ref{eq:ME2} and Eq. \ref{eq:Variance!} a value chemical potential $\mu$ corresponding to the anterior of the embryo, for different $J$; vertical bars indicate $2\tau_c$. We note that we shifted the value of $\varepsilon_b$ for these simulations, as shown in panel $D$. C) Standard deviation as a function of occupation $\bar{C}$ shows characteristic $\bar{C}(1-\bar{C})$ shape and increases with clustering strength. D) The binding energy $\varepsilon_b$ that guarantees half occupation at position $x\approx 0.47$ decreases almost linearly as a function of clustering strength $J$. E) Hill function fits to the mean occupation $\bar{C}$ show that the effective Hill coefficient $h$ approaches the 6 binding sites from below but can exceed this number for an extended sensor with 10 sites. F) The correlation time increases with clustering strengths throughout different values of chemical potential along the embryo. \label{fig:ts}}
\end{figure*}

\begin{equation} \label{eq:C1}
    \bar{C}(\tau) = \dfrac{1}{\tau} \sum_{i=1}^{2^N} \tilde{c}_{i} \left( \sum_{k=1}^{2^N} U_{ik} A_{kk} \sum_{j=1}^{2^N} (U^{-1})_{kj}  \tilde{P_j}(0)\right) \textrm{.} 
\end{equation}
For small systems of $N=10$ and $N=6$ sites,  Eq. \ref{eq:C1} can be solved semi-analytically: we can obtain an exact analytic expression that is too complicated to write down. We plot this solution for the mean occupation from this semi-analytic treatment in Fig. \ref{fig:2}B in our main manuscript. We note that the mean occupations for two different $J$ in Fig. \ref{fig:2}B are calculated with a different value for $\varepsilon_b$, so that the occupation mean for higher $J$ shows its half occupation at $x=0.47$. We choose the value of $\varepsilon_b$ such that this is the case for every $J$ (Fig. \ref{fig:ts}D). Fitting Hill functions to the means obtained with Eq. \ref{eq:C1} for these parameter combinations shows that for increasing $J$ the fitted Hill coefficient approaches the number of binding sites from below ($h \to N=6$) for the binding site region (see Fig. \ref{fig:ts}E); for the extended DNA sensor, we see that for high clustering strengths, higher Hill coefficients can be achieved.

We calculate the variance using Eq. \ref{eq:var2_1}. 
As for the mean, we can obtain a semi-analytic solution by solving the master equation, analogously to Eq. \ref{eq:C1}. Reference \cite{SkogeWingreen} obtained an analytic expression for the variance for the system with $N=6$ binding sites under periodic boundary conditions. We show the agreement of this limit and the semi-analytic solution for the variance from our master equation treatment in Fig. \ref{fig:ts}A,B, and discuss in the next paragraphs how to obtain this limit for our system.

We follow Refs \cite{SkogeWingreen} and \cite{ZhangBerry}, noting that a symmetric transition matrix, $\tilde{W}$, which shares eigenvalues with $M$, can be obtained from $M$ via 
\begin{equation}
\tilde{W}_{ij} = (\tilde{P}_{j}(0)/\tilde{P}_{i}(0))^{1/2} M_{ij}\textrm{,}
\end{equation}
where $\tilde{P}_{i}(0)$ corresponds to the probability of state $i$ at equilibrium. The eigenvectors $\tilde{v}^{(k)}$ of $\tilde{W}$ are orthonormal, and can be related to the eigenvectors $v^{(k)}$ of $M$: $v_i^{(k)}=\sqrt{\tilde{P}_i(0)}\tilde{v}_i^{(k)}$\cite{ZhangBerry}. This implies that the elements of $U^{-1}$ can be expressed in terms of the elements of $U$ via $(U^{-1})_{ki}\tilde{P}_i(0)=U_{ik}$. With that, the probabilities $\tilde{P}_i(t)$ can be fully expressed in terms of the eigenvectors and eigenvalues of $M$. Using $\delta \tilde {c} = \tilde {c} - \langle \tilde{c} \rangle$ to denote deviations from the occupation at steady state, we can then write the variance starting from Eq. \ref{eq:var2_1}
\begin{align}
\sigma_{C}^2 & 
= \frac{2}{\tau^2} \int_{0}^{\tau} dt \kappa(t) (\tau-t) \textrm{,}\nonumber\\
&= \frac{2}{\tau^2} \int_{0}^{\tau}dt \sum_{i=1}^{2^N} \delta \tilde{c}_{i}  \left( \sum_{k=1}^{2^N} v^{(k)}_{i} e^{\lambda_k t} \sum_{j=1}^{2^N}v^{(k)}_{j} \delta \tilde{c}_{j} \right) (\tau-t)\textrm{,}\nonumber\\
&= \frac{2}{\tau^2} \sum_{k=1}^{2^N} \Big( \sum_{i=1}^{2^N} \delta \tilde{c}_{i} v^{(k)}_{i} \Big) \Big( \sum_{j=1}^{2^N} \delta \tilde{c}_{j}v^{(k)}_{j} \Big) \int_{0}^{\tau}dt e^{\lambda_k t}  (\tau-t)\textrm{,}\nonumber\\
&=  2 \sum_{k=2}^{2^N} \Big( \sum_{i=1}^{2^N} \delta \tilde{c}_{i} v^{(k)}_{i} \Big)^2 \Big( \frac{e^{\lambda_k \tau} - \lambda_k \tau -1 }{\tau^2 \lambda_k^2} \Big) \textrm{,}
\end{align}
where the terms $\sum_{i=1}^{2^N} \delta \tilde{c}_{i} v^{(k)}_{i} $ correspond to a projection of the fluctuations in occupation along the $k$th eigenvector and the sum in the last line starts with the second eigenvector, as $\sum_{i=1}^{2^N} \delta \tilde{c}_{i} v^{(0)}_{i} = 0$. 
Finally, the instantaneous variance at equilibrium corresponds to the correlation function $\kappa(0)$ and can be expressed as 
 \begin{align}
\sigma_{\textrm{inst}}^2 = \sum_{k=2}^{2^N} \Big( \sum_{i=1}^{2^N} \delta \tilde{c}_{i} v^{(k)}_{i} \Big)^2 \textrm{.}
\end{align}
While in principle all eigenvalues contribute to the sum, for long averaging times, we can only consider the contribution from the largest negative eigenvalue. Similarly, while the correlation time 
\begin{align} \label{eq: correlation time definition}
\tau_c = \frac{1}{\sigma_{\textrm{inst}}^2} \int_{0}^{\infty}\kappa(t) dt
\end{align}
in principle involves multiple timescales, for many parameter values it is a valid approximation to consider only the largest of the non-zero (negative) eigenvalues, meaning that $\tau_c \approx 1/\lambda_2$. 
Therefore, following Ref \cite{SkogeWingreen}, we find that 
\begin{align}\label{eq:Variance!}
\sigma_{C}^2 \approx \sigma_{\textrm{inst}}^2 \mathcal T (\tau/\tau_c) \textrm{,}
\end{align}
with 
\begin{align}
\mathcal T (x) = 2 \frac{e^{-x} + x -1 }{x^2}\textrm{.}
\end{align}

We show the match of this expression to our semi-analytical calculation from master equation in Fig. \ref{fig:ts}A,B, as a function of measurement time $\tau$ for different $J$ and for a constant value of $\mu$. The variance follows the expected behaviour from Eq. \ref{eq:Variance!} for all $J$, with a clear switch from the instantaneous variance to the scale where it decays as approximately $1/\sqrt{\tau}$ at approximately $\tau \approx 2 \tau_c$. 
%The instantaneous variance (at a fixed value of $\mu$) increases with $J$. Thus, clustering increases the variance throughout the embryo, as expected from earlier work on cooperativity. %Yet, clustering also changes the shape of the mean occupation $\bar{C}$. It is the tradeoff between these effects that contributes to the optimality.
We show the match of this semi-analytical calculation and simulations in the main manuscript, for a single value of $\mu$. 

In Fig. \ref{fig:ts}C, we show the variance for several values of $J$ across different mean occupations $\bar{C}$, which illustrates clearly the increase of the variance with $J$, due to the increase in the correlation time $\tau_c$ with $J$, in agreement with \cite{SkogeWingreen}. 

Finally, we show the correlation time from Eq. \ref{eq: correlation time definition}
from the numerical diagonalization of the master equation for different values of chemical potential as a function of clustering strength $J$ in \ref{fig:ts}F. In agreement with previous calculations, the correlation time increases with clustering strength. 

\begin{figure}
\centerline{\includegraphics[width=0.4\textwidth]{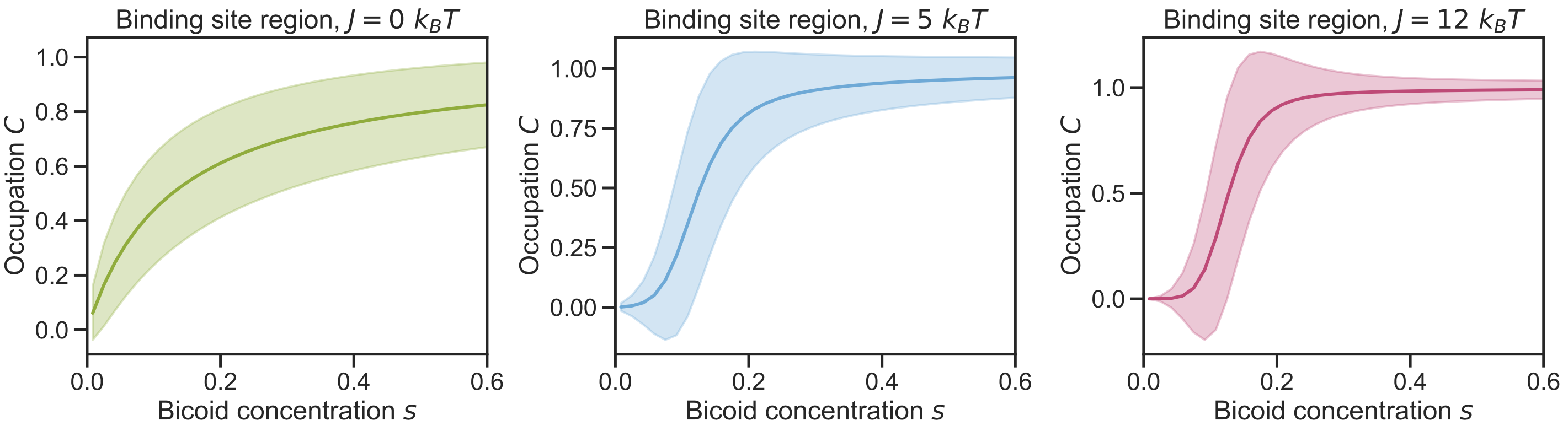}}
    \caption{
Mean and standard deviation change with $J$ in a way that presents a tradeoff for optimizing $I(C;x)$ for the instantaneous measurement with optimum $J \approx 5$. \label{fig:means}}
\end{figure}

In the main manuscript, we discuss the robust optimum at finite clustering strengths for short $\tau$. In Fig. \ref{fig:means}, we show three mean activation profiles to at values of $J$ below, at, and beyond the optimum value of $I(C;x)$ at $J \approx 5$, to illustrate that the instantaneous variance is clearly in a regime where we are not yet in a large-noise limit, and where therefore an infinitely steep threshold is not optimal. 

\subsection{Concentration and occupation sampling for estimating $P(C|s)$ using the mechanistic model \label{sec:appD}}
We model presence and absence of transcription factors with 1 and 0s to avoid symmetry between the presence and absence of transcription factors.  

To estimate the conditional probabilties $P(C|s)$ from simulations of the mechanistic model for clustering transcription factors, both the Bicoid concentration and the occupation must be discretized. To this end, we divide the concentration and occupation ranges into bins, and represent each bin using the value from the center of that bin.

The range of mean Bicoid concentrations (0-147nM) obtained from experimental data (Fig. \ref{fig0}A) was divided into 60 bins. Five additional bins were added, covering higher concentrations, to capture the fluctuations around the maximum mean concentration, resulting in a total of 65 bins covering a range of concentrations from 0-159nM. This number of bins was chosen such that the mutual information between the
position and concentration differed by less than 0.01 from it's limit as the number of bis goes to infinity (as estimated by the mutual information obtained with
$10^5$ bins). The analytical calculations were repeated with a higher concentration sampling using ten times as many bins, to confirm that the choice of concentration sampling does not affect our conclusions (data not shown).

The number of occupation bins chosen for the histogram used to estimate $P(C|s)$ (Fig. \ref{fig:2}D) influences the results obtained for the mutual information. When the number of bins is too low, the true distribution will not be accurately represented by the histogram, but if the number of bins is too high, the discrete nature of the simulations results will lead to artifacts in the final distribution. For this reason it is important to choose the right number of bins.

To this end, we compared the mutual information obtained using the histogram with that obtained using the Gaussian approximation for different numbers of bins. Since the Gaussian approximation is continuous, the mutual information results obtained using this approximation will  converge to the true values as the number of bins increases. By comparing this to the mutual information calculated using the histogram, we can distinguish changes resulting from the mutual information converging to the true value from artifacts caused by the discrete nature of the simulation results. 

Using this method we chose to use 400 bins for the histogram. For the Gaussian approximation, which does not suffer from the same type of artifacts we used 1000 bins. In the case of all-or-nothing expression described in Appendix \ref{all-or-nothing} 
100 bins were used.

To obtain a value of signal concentration $s$ from the chemical potential $\mu$, we calculate the average concentration on a free, non-binding site. This sets the values of chemical potential that we use along the embryonal axis, $x$, allowing us to simulate the binding site sensor essentially in a grand-canonical ensemble. We use 160 positions between 0.1 and 0.9 embryo length. At each position $x$, the distribution of concentrations is assumed to be Gaussian; the mean concentration at each $x$ is found by assuming that the maximum mean concentration $s=1$ corresponds to 147 nM  \cite{dostatniBicoid1}.

\subsection{IB Optimization \label{sec:appE}}
We find the optimal information bound following \cite{tpb00}: we iteratively optimize $P(C|s)$ from a random starting distribution, using 
\begin{equation}
P(C|s) = \frac{P(C)}{Z(s, \lambda)} \exp\left[ - \frac{1}{\lambda} \int dx P(x|s) \ln \left(\frac{ P(x|s)}{ P(x|C)}\right) \right] ,
\end{equation}
where $Z$ is a normalization constant that ensures that $\int P(C|s) dC=1$. For each value of the Lagrangian multiplier $\lambda$, we iterate until $P(C|s)$ converges; from this optimal $P(C|s)$ we can then calculate a particular point along the bound in the information plane. The optimal $P(C|s)$ is thus determined from the data distribution $P(s,x)$, without any physical or mechanistic constraints on $P(C|s)$. $I(C;s)$ and $I(C;x)$ can be calculated from the optimal $P(C|s)$. Practically, we perform a discrete sum instead of an integral. We also used an approximate expression for the information bound \cite{BauerBialek}, in which both $I(C;x)$ and $I(C;s)$ can be expressed using only a changing Lagrangian multiplier $\lambda$ and $I(s;x)$ (not shown). While it captures the bound approximately, it is slightly below the optimal bound and therefore also slightly below some binding site sensors, making it confusing for the purpose of this manuscript.

\begin{figure}
\centerline{\includegraphics[width=0.4\textwidth]{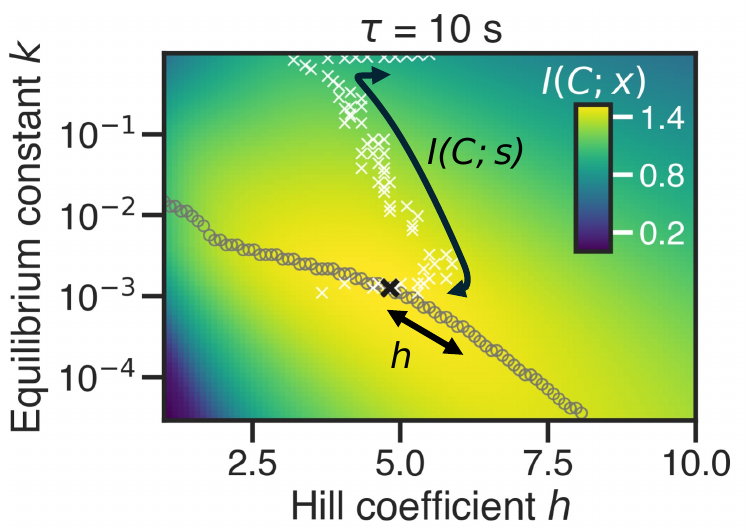}}
\caption{The information $I(C;x)$ as a function of binding site parameters $h$ and $k$ with symbols marking trajectories for a given constraint, for shorter $\tau=10$  s compared to the main manuscript Fig 4B. If $h$ are constrained to varying values of $h$, the optimal $I(C;x)$ lies along the circles, with lower $k$ for higher $h$. If $I(C;s)$ is constrained, the optimal $I(C;x)$ is marked with white crosses. The path of white crossed begins beyond the optimum (see blue line in Fig. 4C, which decreases as $I(C;s)$ is pushed beyond 2 bits). The maximum $I(C;x)$ is marked with a black cross. \label{fig:suppD}}
\end{figure}
To find the binding site sensors that also lie on the optimal bound, we search our sensors numerically: we first find all sensors with a particular $I(C;s)$ and then choose the sensor with the highest $I(C;x)$. This yields the white crosses in Fig. 4B and Fig. \ref{fig:suppD} (for $\tau =600$ s and $\tau =10$ s, respectively). In Fig. \ref{fig:suppD} we see the white path begin at smaller $h$ than the optimum, which corresponds to the part of the optimal bound in Fig 4C where $I(C;x)$ decreases for $I(C;s)>2$. As $I(C;x)$ decreases from its largest possible value, the path in the binding site parameter plane moves away from the optimum, towards larger values of $k$ and towards the boundary of the parameters we sample. When $I(C;s)$ just decreases away from the maximum, $h$ increases, but then decreases as the $I(C;s)$ decreases further. Eventually, the path arrives at the maximal value for $k$ we consider; if we increase $k$ further, the path continues along its current trajectory. However, in this case the binding site sensors along the optimal bound yield mean occupations that no longer include full occupation of the binding site at high $s$. Since we believe that such binding sites are no longer sensible, we typically only consider binding sites with a maximal normalized occupation $\bar{C}=0.6$; if we do so, the path continues towards significantly higher $h$ for almost constant $k$. This behaviour is again in principle consistent with the idea that lower $I(C;s)$ implies higher $h$, though we do note that we have effectively imposed an additional constraint on the binding site occupation; without this constraint, $h$ would continue to decrease slightly and $k$ would increase significantly, which implies an unphysically small $k_\textrm{on}$.

We hope to show with this trajectory in Figs. 4B and \ref{fig:suppD}  that the constraints imposed matter significantly for finding the optimal binding site parameters.

\subsection{All-or-nothing expression
\label{all-or-nothing}}
\begin{figure}
%  \centerline{\includegraphics[width=.49\textwidth]{Manuscript_MutualInformationSingleMeasurement_AllOccupiedExpression.png}}
% \centerline{\includegraphics[width=.49\textwidth]{Manuscript_MutualInformationRepeatedMeasurement_AllOccupiedExpression.png}}
\centerline{\includegraphics[width=.49\textwidth]{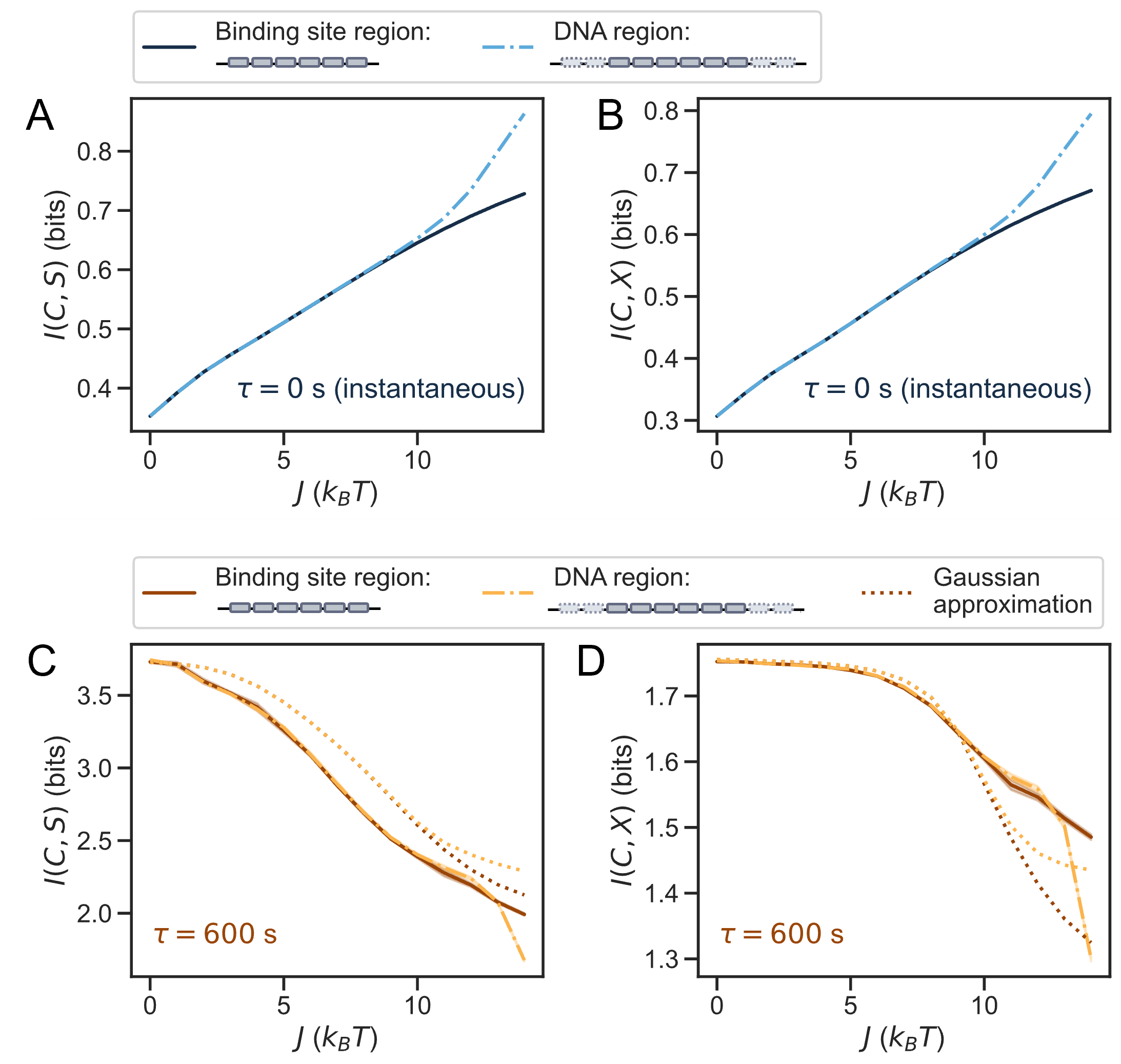}}
\caption{ Mutual information for a transcription factor measurement which is processed in 'all-or-nothing' fashion, i.e., where occupation is read out only in terms of two states: one state corresponding to a fully occupied binding site, that can activate gene expression downstream, and another state representing all other occupation states. \label{fig:8}}
\end{figure}

We assumed for almost all of the manuscript that gene expression regulated by the binding site sensor is proportional to the averaged occupation at the sensor. This seems biologically relevant, as polymerases could be recruited proportional to this average occupation, though of course this recruitment would add further noise into the process. Nevertheless, recent work on the hbP2 enhancer also discusses a scenario where gene expression (or polymerase recruitment) can only occur when the binding site is fully occupied \cite{Desponds2020}. 

We study this scenario numerically. We analyse our simulated trajectories for the long measurement time and instantaneous measurement, to obtain mean and variance for such a measurement $C'$. These can be used to calculate probability distributions $P(C'|s)$ as before. The mutual information $I(C';s)$ and $I(C';x)$ is shown in Fig. \ref{fig:8}. For all measurements, the values of information are slightly lower than for the measurement that reads out the time-averaged occupation, since the all-or-nothing readout presents a course-graining that reduces information, consistent with the data-processing inequality.  For the instantaneous measurement, clustering increases both relevant information quantities for all clustering strengths studied here; for the longer measurement, both informations decrease, with a strong plateau for $I(C';x)$ for weak clustering strengths as in the main manuscript. Therefore, we conclude that our result that weak clustering is information-theoretically acceptable applies also to this different readout protocol. 
\bibliography{references_text}

\end{document}